\documentclass[a4paper,11pt]{article}
\pdfoutput=1

\usepackage{jheppub} %
\usepackage{graphicx}
\usepackage[english]{babel}
\usepackage{hyperref}                               %
\usepackage[dvipsnames]{xcolor}                       %
\usepackage{cleveref}                               %
\usepackage{float} 									%
\usepackage{xparse}                                 %
\usepackage{bm}                                     %
\usepackage{amsmath}                                %
\usepackage{tikz}                                   %
\usepackage{tensor}                                 %
\usepackage{comment}
\usepackage{placeins}                               %
\usepackage{cancel}

\usepackage{todonotes}
\presetkeys{todonotes}{inline}{}

\renewenvironment{align}{
    \begin{equation}
    \begin{aligned}
}{
    \end{aligned}
    \end{equation}
    \ignorespacesafterend
}

\crefname{figure}{Figure}{Figure}
\crefname{equation}{}{}
\crefname{table}{Table}{Table}
\crefname{section}{Section}{Section}
\crefname{appendix}{Appendix}{Appendix}

\hypersetup{
    colorlinks,
    linkcolor={blue},
    citecolor={blue},
    urlcolor={blue}
}

\usepackage{parskip}				%
\renewcommand*{\d}{\, \mathrm{d}}      %

\newcommand{\lp}{\left(}
\newcommand{\rp}{\right)}
\newcommand{\lsb}{\left[}
\newcommand{\rsb}{\right]}

\renewcommand{\Re}{\operatorname{Re}}
\renewcommand{\Im}{\operatorname{Im}}

\newcommand{\abs}[1]{\left| #1 \right|}
\NewDocumentCommand\dv{m+g}{%
  \IfNoValueTF{#2}
    {\frac{\d}{\d {#1}}}
    {\frac{\d {#1}}{\d {#2}}}%
}
\NewDocumentCommand\pdv{m+g}{%
  \IfNoValueTF{#2}
    {\frac{\partial}{\partial {#1}}}
    {\frac{\partial {#1}}{\partial {#2}}}%
}
\NewDocumentCommand\vdv{m+g}{%
  \IfNoValueTF{#2}
    {\frac{\delta}{\delta {#1}}}
    {\frac{\delta {#1}}{\delta {#2}}}%
}

\makeatletter
\newcommand*{\eval}[1][\bigg]{%
   \if\relax\detokenize{#1}\relax
      \def\next{\mathclose|}%
   \else
      \def\next{\csname\expandafter\@gobble\string#1r\endcsname|}%
   \fi
   \next
}
\makeatother

\date{\today}

\title{Quantum critical theories in a periodic potential: strange metallic thermoelectric and magnetotransport}

\abstract{
We study DC and AC thermoelectric and magneto-transport in 2D quantum critical theories
with strong translational symmetry breaking due to a %
varying
chemical potential lattice with zero average $\bar{\mu}=0$. 
The combination of quantum criticality and the absence of the average natural scale implies that such systems have idiosyncratic signatures that may apply more generally when the variance in the lattice potential far exceeds the average or for strong translational symmetry breaking in general. 
We model such theories holographically through near-extremal AdS black holes. 
We find that 
these systems (a) become {\em better} conductors.
In a 2D lattice, this can be explained by currents flowing around obstacles;
(b) exhibit bad-metal electrical transport with Drude-like thermal transport, 
though it is not Drude, and,
notably, (c)
display 
an approximately
$B$-linear longitudinal magnetoresistance at large fields, similar to Effective Medium Theory.
We 
comment on how these results may apply when $\bar{\mu}\neq 0$.

}

\begin{document}
\author[a]{E.~Nilsson,}
\author[b]{K.~Schalm}

\affiliation[a]{Department of Physics,
Division of Subatomic, High Energy and Plasma Physics,
Chalmers University of Technology\\
SE-412 96 G\"{o}teborg,
Sweden}
\affiliation[b]{Instituut-Lorentz, $\Delta$ITP, Universiteit Leiden, P.O. Box 9506, 2300 RA Leiden, The
Netherlands}

\emailAdd{nieric@chalmers.se}
\emailAdd{kschalm@lorentz.leidenuniv.nl}

\maketitle
\flushbottom

\section{Introduction}
\label{sec:introduction}
Essentially all critical points studied in real materials require fine tuning to a very special point among all of its possible configurations. 
Nevertheless, this special point is of fundamental physical interest, due to universality:
the physics at a critical point is independent of its microscopic origins. The particularities of the material in question no longer matter as long as one is able to tune to the special point. If the material is a (crystalline) solid, the crystal lattice is generically such a particularity. In the formal sense, the lattice is irrelevant under the renormalization group flow to the critical point. Precisely because of this fact, a lattice is often used in numerical analyses as a regulator.

Quantum critical theories, where the critical point exists at $T=0$, have the added factor that the critical behavior extends fan-wise into the phase diagram at finite temperature $T$ \cite{sachdev_quantum_2023}. The temperature $T$ acts as an IR regulator, and probes of the system at an energy scale less than $T$ are controlled by collective hydrodynamic rather than pure scale-invariant and collisionless behavior. Importantly, the entrance to the hydrodynamic regime halts the RG flow, and whatever remains of the lattice is imprinted in the response. A striking example is a weak chemical potential lattice, $\mu(x)=\bar{\mu}+\delta \mu \cos(Gx)$, with lattice momentum $G < \bar{\mu}$ \cite{balm_t-linear_2023}: in this case the IR is described by Umklapp hydrodynamics based on Bloch waves, rather than conventional hydrodynamics based on Fourier modes. %
When instead $G > \bar{\mu}$, the conventional notion that the lattice is irrelevant holds. 

A chemical potential lattice also has a third scale $\delta\mu$, encoding the size of the lattice variations.
Intuited by real metals with a finite density of valence electrons, one normally studies the weak lattice regime, where the average chemical potential is parametrically larger than the size of these variations; $\bar{\mu}\gg \delta \mu$. For quantum critical theories placed in such a lattice, all changes w.r.t.~the translational invariant theory are then encoded as functions of two dimensionless parameters $\bar{\mu}/G$ and $\delta \mu/G$.

Chesler, Lucas, and Sachdev realized that 
the study of
the strong lattice regime $\delta\mu \gg \bar{\mu}$ in such critical theories becomes tractably simpler by taking the extremal value $\bar{\mu}=0$ \cite{chesler_conformal_2014}. This eliminates a background scale, simplifying the structure of the solutions and opening the possibility of universal transport regimes exhibiting scaling collapse. This may therefore serve as a testing ground for strongly correlated physics near or at quantum criticality where translational symmetry breaking is strong.
In this article we extend their findings of electronic transport in this special regime %
from 1D to 2D lattices, covering DC and AC thermal, thermoelectric, and magnetotransport.
As Chesler, Lucas and Sachdev, we use holographic models of quantum criticality to compute these responses.

For an experimental realization of such physics, graphene at charge neutrality naturally comes to mind. Indeed, many studies have considered 
the scenario 
of a spatially periodic chemical potential imposed on $\bar{\mu}=0$ charge neutral graphene either in terms of large scale ripples (a 1D periodic potential) or an egg-carton landscape (a 2D periodic potential), see e.g.,~\cite{parkAnisotropicBehaviorsMassless2008,hoSemimetalicGrapheneModulated2009,parkNewGenerationMassless2008,wangElectronicBandGaps2010,bursetTransportSuperlatticesSingle2011,yankowitzEmergenceSuperlatticeDirac2012,breyNonlocalQuantumEffects2020,fernandesEffectiveMediumModel2023}. Most of these adopt a non-interacting or weakly interacting point of view, even though charged excitations in graphene are arguably not weakly coupled. For 1D potentials, these weak coupling results  
agree qualitatively with the strong coupling approach in \cite{chesler_conformal_2014}. %
We shall show that this is not so for a 2D periodic potential, where the weak coupling approach surprisingly predicts no change to electrical transport \cite{bursetTransportSuperlatticesSingle2011}. Our approach at strong coupling does however show distinct physics that also differs from the 1D case; thus, both interactions and dimensionality matter.

Due to its 
quantum critical-like physics at charge neutrality, graphene is in fact an ``almost strange metal'' \cite{sachdevQuantumCriticalityBlack2009,mullerGrapheneNearlyPerfect2009}.
Our findings are particularly relevant 
for the truly strange metal phase in high-$T_c$ cuprate superconductors 
in light of the 
hypothesis that 
its phenomenology 
is 
described by such quantum critical physics in the presence of a 2D lattice; see e.g., \cite{zaanen_holographic_2015,hartnoll_holographic_2018} for a review. 
This was recently exemplified by computations in a 2D Yukawa-SYK model, which has a holographic dual. This model 
reproduces the the observed linear-in-$T$ DC conductivity, an AC conductivity with apparent $\omega^{-2/3}$ mid-IR scaling, $T\ln T$ specific heat, Homes' law relating the $T=0$ superfluid density to $T_c\sigma_{DC}(T_c)$, and broad ARPES width from marginal Fermi liquid self-energies $\Sigma(\omega) \sim \omega\ln\omega$ \cite{li_strange_2024}.
The presence of translational symmetry breaking --- in \cite{li_strange_2024} mainly through disorder in addition to the lattice --- is paramount in order to obtain finite and properly scaling transport properties.
However, one important phenomenological observation is hitherto unexplained: the magnetotransport. No models as of yet are able to convincingly explain the observed Hall angle scaling ${\frac{\sigma_{xx}}{\sigma_{xy}}\sim T^2}$.
In systems dominated by a single momentum relaxation rate, this rate determines both the Hall angle and DC resistivity and these must scale the same way with temperature. %
Anderson already pointed out that to explain this observed difference in the cuprate strange metal one needs a second relaxation rate, which is not trivial \cite{andersonHallEffectTwodimensional1991a}. As one of us has recently shown \cite{chagnet_natural_2024}, %
extending Anderson's earlier arguments to a full hydrodynamic analysis %
requires a system in which momentum relaxation through lattice Umklapp or disorder is \emph{not} the only parametrically small relaxation rate, independently of whether the microscopics is strongly or weakly coupled.
An important ingredient in reaching this conclusion is the AC response, in addition to the DC response. While analyses based solely on DC transport have been remarkably successful in describing strange metals \cite{amorettiHydrodynamicalDescriptionMagnetotransport2020,blakeQuantumCriticalTransport2015}, such theories typically rely on a separation of mechanisms: the longitudinal DC resistivity is assumed to be of quantum critical origin, whereas the Hall conductivity is not. This effectively corresponds to scenarios in which the system behaves as if it were charge neutral, or exhibits strong translational symmetry breaking for longitudinal transport but weak symmetry breaking at finite charge density for Hall transport. This picture is prima facie in tension with measurements of the AC longitudinal optical conductivity, which display a clear Drude peak at low temperatures. 
The search for systems that exhibits both Drude-like AC optical conductivity and anomalous Hall transport therefore motivates our study of systems in the extreme limit of strong translational symmetry breaking.

Experiments on certain
cuprates (and pnictides) also reveal a longitudinal magnetoresistance on the form \cite{hayes_scaling_2016,ayres_incoherent_2021,giraldo-gallo_scale-invariant_2018}
\begin{equation}
    \rho(B,T)-\rho(0,0) \propto 
    \sqrt{(\alpha k_{B}T)^{2}+ (\gamma \mu_{B}B)^2}
    \label{eq:long-magneto-resis}
\end{equation}
where $\alpha$ and $\gamma$ are two dimensionless parameters.
This leads to 
an unsaturating
$B$-linear magnetoresistance at large magnetic fields, which stands similarly in stark contrast to the $B^2$-scaling of conventional Fermi liquids that saturates to a constant at large fields \cite{kittel_introduction_2005}.
$B$-linear magnetoresistance has been shown to be a general feature of systems with large 
spatial
fluctuations in the conductivity, extending down to moderate magnetic fields in the case of strong inhomogeneities \cite{parish_non-saturating_2003}. In that case, such systems can be modeled as random resistor networks, which has been shown to be equivalent to ``Effective Medium Theory'' (EMT) \cite{ramakrishnan_equivalence_2017}. Within EMT, one considers the macroscopic response of a system constituting of multiple subdomains with varying conductive properties,
and may be viewed as an
extreme form of translational symmetry breaking. The case of two different types of subdomains at an equal area fraction yields an exactly $B$-linear magnetoresistance \cite{guttal_model_2005}.\footnote{In holographic approaches there exist so-called DBI models which predict a resistivity of the form \cref{eq:long-magneto-resis} \cite{kiritsisQuantumCriticalityDBI2017}, but the physical origin is an underlying relativistic symmetry rather than an averaging.}
Whether all strange metals exhibit a magnetoresistance on quadrature form as in \cref{eq:long-magneto-resis} is questioned \cite{boyd_single-parameter_2019}; nevertheless, a roughly $B$-linear, unsaturating magnetoresistance appears to be a robust feature.

Both of these magnetotransport observations suggest that systems with strong translational symmetry breaking --- whether due to an underlying lattice or disorder --- should be considered to explain the phenomenology of cuprate strange metals. This provides an additional motivation to study 
quantum-critical systems in the charge-neutral regime, where the spatially averaged chemical potential vanishes ($\bar{\mu}=0$) but fluctuates locally, $\delta\mu(\vec{x})\neq 0$, beyond purely the fundamental theoretical interest in its universal response. 
Moreover, at charge neutrality, thermal and electrical transport naturally decouple. In such regimes there are therefore two different independent transport mechanisms that are important in the IR. The resulting universal features are most naturally captured within a holographic framework, which generalizes Landau–Ginzburg theory rather than relying on a specific SYK-like model \cite{zaanen_holographic_2015}.
This is the analysis we pursue in this paper. We provide a complete analysis of the thermoelectric response, both AC and DC, numerically computed, for quantum critical conformal field theories 
subject to a spatially modulating chemical potential with zero average. In particular, we consider both 1D (\cref{sec:1Dresults}) and 2D (\cref{sec:2Dresults}) ``lattices'' to find that the dimension of the lattice plays an important role, altering the IR fixed point of the theory, even though both are ultra-local theories evading any RG-like $c$-theorem as found by \cite{chesler_conformal_2014}. %
We find that electric transport is completely incoherent, whereas the thermal transport is predominantly coherent and Drude-like. Turning on a magnetic field, we find preliminary evidence of $B$-linear magnetotransport at large fields. 
Our setup naturally suggests connections to Effective Medium Theory descriptions, and in \cref{sec:conclusion} we conclude with an outlook how a combination of this simplified regime with one dimensionless parameter voided provides an obvious direction for further study.

\section{Review: holographic model of strongly correlated quantum criticality}
\label{sec:holomodel}
To model a generic (large $N$) 2D strongly quantum critical system, we use its holographic dual description as dynamical gravity in 4D asymptotically anti-de-Sitter space.\footnote{These quantum critical theories have a relativistic Lorentz invariance in the UV. This is the Lorentz invariance extended from quantum critical excitations with linear dispersion, where the emergent velocity $\omega/k$ plays the role of the speed of light. For most observables this Lorentz symmetry is broken in the IR.} Those familiar with this approach may skip directly to \cref{sec:hydro_exp}.
Including a Maxwell field dual to the global $U(1)$ of electromagnetism,
we consider the AdS$_4$ 
action
\begin{equation}
     S = \frac{1}{2\kappa^2} \int \d^4x \, \sqrt{-g} \lsb R -  2 \Lambda - \frac{1}{4}F_{\mu\nu}F^{\mu\nu}\rsb\,,
     \label{eq:S}
\end{equation}
with ${\Lambda = -3/L^2}$ the negative cosmological constant; %
we set the AdS length scale $L=1$ without loss of generality. The equations of motion in the bulk are the (trace-reversed) Einstein's equations and Maxwell's equations,
\begin{align}
    R^H_{\mu\nu} &= \kappa^2 \lp T_{\mu\nu}- \frac{1}{2}T g_{\mu\nu}\rp\,, \quad &T_{\mu\nu} &= \tensor{F}{_\mu^\rho}F_{\nu \rho} - g_{\mu \nu} \lp \frac{1}{4}  F_{\rho \sigma}F^{\rho \sigma} + 2 \Lambda \rp\, , \\
    \nabla_\mu F^{\mu\nu} + \nabla^\nu \varphi &= 0\,, \quad &\varphi &= \nabla_\mu (A^\mu - \bar{A}^\mu) + \xi^\mu (A_\mu - \bar{A}_\mu)\, .
    \label{eq:bulk_EOM}
\end{align}
We employ the de Turck trick to make the equations elliptic and the boundary value problem well-defined by substituting the Ricci tensor with the Harmonic Ricci tensor: ${R^H_{\mu\nu} = R_{\mu\nu} - \nabla_{(\mu} \xi_{\nu)}}$ \cite{headrick_new_2010}. This imposes the gauge ${\xi^\mu = 0}$, where ${\xi^\mu = g^{\lambda\nu}(\Gamma^{\mu}_{\lambda\nu}-\tilde{\Gamma}^{\mu}_{\lambda\nu})}$ is the de Turck vector and $\tilde{\Gamma}^{\mu}_{\nu\rho}$ is the Christoffel symbol for a suitably chosen background reference metric, which we take to be the analytical solution for an isotropic system \mbox{($\delta\mu = 0$)}.
In the absence of a background magnetic field, this reference geometry is 
the 
AdS$_4$-Reissner-Nordstr{\"o}m
solution, whereas for finite magnetic fields we consider the 
dyonic Reissner-Nordstr{\"o}m black brane. We also add a similar gauge-fixing term to Maxwell's equations in \cref{eq:bulk_EOM}, enforcing $\varphi=0$, where ${2 \bar{A} = B (x \d y - y \d x)}$ \cite{donos_minimally_2016,donos_holographic_2020}. This ensures that one can consistently periodically identify the spatial directions $x$ and $y$ as is necessary for the numerics, and imposes a modified Lorenz gauge for the periodic part $A_{\mu}-\bar{A}_{\mu}$ of the background gauge field.

We solve these equations numerically. A well-converged numerical solution to the modified equations \cref{eq:bulk_EOM} leads to a simultaneous solution of the Einstein-Maxwell equations and the gauge constraints. We verify the convergence of our numerical algorithm by an a posteriori check that $\varphi$ and $\xi^2$ are numerically close to zero, as is detailed in \cref{app:numerics}.
In the most general case of a 2D lattice and a finite magnetic field $B$, our ansatz for the background fields is
\begin{align}
    \d s^2 &= \frac{L^2}{z^2} \lsb - f(z) Q_{tt}\eta_t^2 +  Q_{xx}\eta_x^2 + Q_{yy} \eta_y^2 + \frac{Q_{zz}}{f(z)} \d z^2\rsb\, , \\
     \eta_t &= \d t + Q_{tx} \d t + Q_{ty} \d y + Q_{tz} \d z\,, \quad \eta_x = \d x + Q_{xy} \d y + Q_{xz} \d z\,,
     \\ \eta_y &= \d y + Q_{yz} \d z\,, \quad f(z) = (1-z)\lp 1 + z + z^2 - \frac{\bar{\mu}^2+B^2}{4}z^3\rp\,, \\
    A &= (1-z) a_t \d t + \lp a_x - \frac{B}{2} \rp \d x + \lp a_y + \frac{B}{2} \rp \d y + a_z \d z\,,
    \label{eq:bg_ansatz}
\end{align}
where $Q_{\mu\nu}=Q_{\mu\nu}(x,y,z)$, $a_\mu = a_\mu(x,y,z)$ and $z$ is an inverted radial coordinate, such that the AdS conformal boundary sits at $z=0$. 
This ansatz describes a dyonic black hole of temperature
\begin{equation}
    T = \frac{12 -\bar{\mu}^2 - B^2}{16 \pi}\,,
    \label{eq:T_BH}
\end{equation}
where we have set the black hole horizon location $z_h = 1$. 
For nearly all the data presented, we limit ourselves to the universal regime where \mbox{$\bar{\mu}=0$}. 
When $B=0$, we reduce the complexity of the ansatz by directly setting ${Q_{ti} = a_{i} = 0}$, where ${i=\{x,y,z\}}$.\footnote{For 1D lattices we also drop the $y$-dependence and set $Q_{xy}=Q_{yz} = 0$.} 
In the UV ($z=0$), we demand that the metric asymptotes to AdS$_4$ and that the time component of the gauge field equals the spatially modulating chemical potential, i.e.,
\begin{equation}
    a_t  = \mu(\vec{x})\,,
    \quad a_i = 0\,, \quad Q_{ti}=Q_{ij}\vert_{i\neq j} = 0, \quad Q_{\nu\nu} = 1 \,,\quad \nu = \{t,x,y,z\}\,.
    \label{eq:uv-bc}
\end{equation}
The IR boundary conditions ($z=z_h$) follow from algebraically solving the equations of motion on the horizon, which leads to the Dirichlet condition $(Q_{tt} - Q_{zz})\vert_{z=z_h} = 0$ and Robin-type boundary conditions for the other fields. The resulting non-linear equations of motion are solved numerically on the domain $x,y \in [0, 2 \pi/G)$, $z \in [0, 1]$ with periodic boundary conditions in the transverse $x$ and $y$ directions. Further details on the numerical implementation can be found in \cref{app:numerics}.

\subsection{Extracting quantum critical responses}

Following Refs.~\cite{donos_thermoelectric_2014,donos_thermoelectric_2015,banks_thermoelectric_2015,donos_holographic_2017}, we obtain the DC thermoelectric transport coefficients in the dual quantum critical theory by solving a Stokes flow problem on the horizon on the black hole for a given bulk geometry. This reduces the number of degrees of freedom needed to be solved for as compared to an AC calculation, drops the degree of the PDE by one, and avoids the need to take the numerical ${\omega \to 0}$ limit of AC data. The Stokes equations read \cite{banks_thermoelectric_2015,balm_t-linear_2023}
\begin{align}
     2 \sqrt{h} \nabla^j \nabla_{(i}v_{j)} + \d \chi_{ij}^{H} Q^j + F_{ij}^{H}J^j + 
    s^{H}T \zeta_i - \sqrt{h}\nabla_i p + 
      n^{H}(E_i + \nabla_i w) = 0\,, \\
    \nabla_i Q^i = 0\,, \qquad \nabla_i J^i = 0\,,
    \label{eq:stokes}
\end{align}
where $i=\{x,y\}$.
These equations, sourced by an electric field $E_i$ and a thermal gradient $\zeta_i$, are to be solved for the four unknowns $v_x,v_y, w$ and $p$, which are functions of $x$ and $y$. The electric and heat currents $J^i$ and $Q^i$ and the transport coefficients depend on these functions as
\begin{align}
    J^i &= n^{H} v^i + \sqrt{h}(E^i + \nabla^i w + F^{ij}_H v_j)\,, \quad Q^i = s^{H}T v^i\,, \\
    s^{H} &= 4 \pi \sqrt{h}\,, \quad n^H = \sqrt{h} \frac{a_t}{Q_{tt}}\vert_{z=z_h}\,, \quad \chi_i^H  = Q_{ti}\vert_{z=z_h}\,, \quad F_{ij}^H = 2\partial_{(i} A_{j)}\vert_{z=z_h} \,,
    \label{eq:stokes_formulae}
\end{align}
where indices are raised and lowered with respect to the horizon metric $h_{ij} = g_{ij}\vert_{z=z_h}$. The key idea of \cite{banks_thermoelectric_2015} is that the spatial averages $\Bar{J}^i$ and $\Bar{Q}^i$ (taken over the unit cell) do not renormalize as one moves from the horizon to the boundary. The DC thermoelectric response functions in the boundary theory may therefore be extracted from
\begin{equation}
    \begin{pmatrix}
        \Bar{J}_i \\
        \Bar{Q}_i
    \end{pmatrix}
    = 
    \begin{pmatrix}
        \sigma_{ij} & T \alpha_{ij} \\
        T \bar{\alpha}_{ij} & T \bar{\kappa}_{ij}
    \end{pmatrix}
    \begin{pmatrix}
        E^j \\
        \zeta^j
    \end{pmatrix}
    \, .
\end{equation}
In our numerics, we fix either the electric field $E^i$ or the thermal gradient ${\zeta^i = - \nabla^i T/T}$ and study the extracted electric and heat currents $J^i$ and $Q^i$. 
In the presence of a nonzero magnetic field $B$, the boundary electric and heat currents decompose into transport and magnetization pieces where the latter should be subtracted in order to obtain the physical transport currents that couple to the external electric field or thermal gradient \cite{hartnoll_theory_2007,blake_magnetothermoelectric_2015}. However, the magnetization currents vanish at the horizon \cite{donos_dc_2016}, and as a result, the horizon currents computed via the Stokes flow method directly yield the physical DC conductivities.
\subsection{Extracting AC transport}
For transport at finite frequency, we consider linear perturbations of the background geometry. In the linearized versions of the bulk equations of motion \cref{eq:bulk_EOM}, the linearized de Turck term reads
\begin{equation}
    \nabla_{(\mu} \tau_{\nu)}\,, \quad \text{where} \quad \tau_{\mu} = \nabla^{\nu}\delta(g_{\mu\nu}) - \frac{1}{2} \nabla_{\mu} \delta({g^{\nu}}_{\nu})\,,
    \label{eq:deDonder}
\end{equation}
which imposes the de Donder gauge $\tau_\mu = 0$. Additionally, we add 
a 
gauge-fixing term to the left-hand side of the Maxwell equations as
\begin{equation}
    \delta(\nabla_\nu \nabla^{\nu} A_\mu - \nabla_\mu \nabla_\mu A^\nu) + \nabla_{\mu}\nabla^\nu \delta (A_{\nu}) = 0\,,
    \label{eq:Lorenz}
\end{equation}
imposing the Lorenz gauge $\nabla^\nu \delta (A_{\nu}) = 0$. This ensures that each of the fields in the ansatz below has a kinetic term (i.e., the equations of motion contain a term $\partial_z^2 \delta \Phi$ for all perturbations $\delta \Phi$).
We make an ansatz for the perturbed metric and gauge field as
\begin{equation}
    \delta (g_{\mu\nu}) = \frac{e^{- i \omega t}}{z^2} \delta g_{\mu\nu}(x,y,z)\,,
    \quad 
    \delta (A_{\mu}) = e^{- i \omega t} \delta A_\mu(x,y,z)\,,
\end{equation}
which in the most general case leads to 14 unknown functions $\{ \delta g_{tt}, \dots, \delta g_{zz}, \delta A_t, \dots \delta A_z \}$ to be solved for.
The IR boundary conditions are obtained by factoring out an infalling factor and algebraically solving the equations of motion at the horizon.
The UV boundary conditions impose either a source for the charge (fixing $\delta A_x$) of for the heat (fixing $\delta g_{tx}$), setting the remaining fields to equal zero. Although the resulting system of equations is linear, it is in general much more  numerically challenging to
solve than the non-linear equations for the background.

We compute the frequency-dependent thermoelectric response from retarded two-point functions of the conserved U(1) current $J^\mu$ and the energy-momentum tensor $T^{\mu\nu}$. The (boundary) heat current is defined by
\begin{equation}
    Q^i = T^{t i}- \mu J^i\,.
\end{equation}
With our conventions, we obtain
\begin{align}
    \sigma^{ij}(\omega) &= \frac{i}{\omega} \Big [ G^{R}_{J^i J^j}(\omega,k=0)- \Re G^{R}_{J^i J^j}(\omega=0,k\to 0) \Big ]\,,\\
    T \alpha^{ij}(\omega) &= \frac{i}{\omega}\lsb G^{R}_{J^i Q^j}(\omega,k=0) - \Re G^{R}_{J^i Q^j}(\omega=0,k\to 0) \rsb\,,\\
    T \bar\kappa^{ij}(\omega) &= \frac{i}{\omega}\lsb G^{R}_{Q^i Q^j}(\omega,k=0)- \Re G^{R}_{Q^i Q^j}(\omega=0,k\to 0) \rsb\,,
\end{align}
and average the results over the unit cell. The subtraction of the zero-frequency real part removes potential contact terms \cite{kovtun_lectures_2012,davison_dissecting_2015}. We restrict the AC analysis to zero background magnetic field, for which magnetization currents vanish and no subtraction is required.

\section{Transport expectations from hydrodynamics}
\label{sec:hydro_exp}
The periodic potential environment in which we wish to study the quantum critical response, is captured in the UV boundary conditions of our ansatz \cref{eq:uv-bc}.
At finite density, i.e., finite average $\bar{\mu}\neq 0$, a simple (relativistic) hydrodynamic analysis extended with a momentum relaxation rate $\Gamma$ parametrically smaller than any other rate ($\bar{\mu}\gg \delta \mu$ in the case of an chemical potential lattice) results in thermo-electric transport that is dominated by a Drude pole at $\omega=-i\Gamma \sim -i(\delta \mu)^2$, such that the AC transport coefficients take the general form
\cite{davison_dissecting_2015,hartnoll_theory_2007}
\begin{align}
    \sigma(\omega) &= \frac{n^2}{\varepsilon + p} \frac{1}{\Gamma - i \omega} + \sigma_Q \,, \\
    \alpha(\omega) &= \frac{n s }{\varepsilon + p} \frac{1}{\Gamma - i \omega} - \frac{\mu}{T}\sigma_Q \,,\\
    \bar{\kappa}(\omega) &= \frac{s^2 T}{\varepsilon + p} \frac{1}{\Gamma - i \omega} + \frac{\mu^2}{T}\sigma_Q\, ,
    \label{eq:hydro_transport}
\end{align}
where $n, s, \varepsilon$ and $p$ denote the charge, entropy, energy and pressure densities, respectively.
We will be studying the system in the opposite limit $\bar{\mu}\ll \delta \mu$, but the generic wisdom that translational symmetry breaking leads to finite DC conductivities should hold.
In the extremal limit of interest to us,
$\bar{\mu} = n = 0$, a simple exercise in hydrodynamics indicates that electric and thermal transport will decouple.
We might therefore expect the electrical conductivity to be governed by the incoherent conductivity $\sigma_Q$. %
Similarly, 
the thermal conductivity 
should remain coherent (i.e., dominated by a single pole). In the absence of magnetic fields, we furthermore expect $\alpha = 0$, confirming
that the mechanisms of electrical and thermal transport are essentially independent. We thus also expect no difference between the thermal conductivity at zero electric current ${\kappa_{ij} = \bar{\kappa}_{ij} - T \bar{\alpha}_{ik} (\sigma^{-1})_{kl}\alpha_{lj}}$ and $\bar{\kappa}_{ij}$. 

At finite magnetic fields, the transport coefficients become tensorial in nature, and one
naturally expects $\sigma_{xy} \sim n/B$, as follows from a combination of parity and charge-conjugation symmetry.
In a charge neutral system with $n=0$, however, no Hall effect will occur, and the magnetic field will only affect longitudinal transport. %
Nevertheless, a finite magnetic field does allow for a Hall thermoelectric conductivity $\alpha_{xy} \sim s/B$ and we will be able to observe a non-zero contribution to the Nernst coefficient $e_N = - (\sigma^{-1}\alpha)_{xy}$, where from charged hydrodynamics one expects \cite{hartnoll_theory_2007}
\begin{equation}
    e_N = \frac{B}{T}\frac{1}{\Gamma}\frac{1}{(\frac{n^2}{\sigma_Q(\epsilon+p)\Gamma}+1)^2+\omega_c^2}\,,
    \label{eq:Nernst_hydro}
\end{equation}
with the cyclotron frequency in natural units equal to $\omega_c=\frac{Bn}{\epsilon+p}$.\footnote{A review of charged hydrodynamics in the presence of weak translational symmetry breaking is given in \cite{amorettiHydrodynamicsDimensionalStrongly2022b}.}
For a charge neutral system this simplifies to a direct measurement of the momentum relaxation rate as \mbox{$e_N=\frac{B}{T\Gamma}$}.

\begin{figure}
    \centering
    \includegraphics{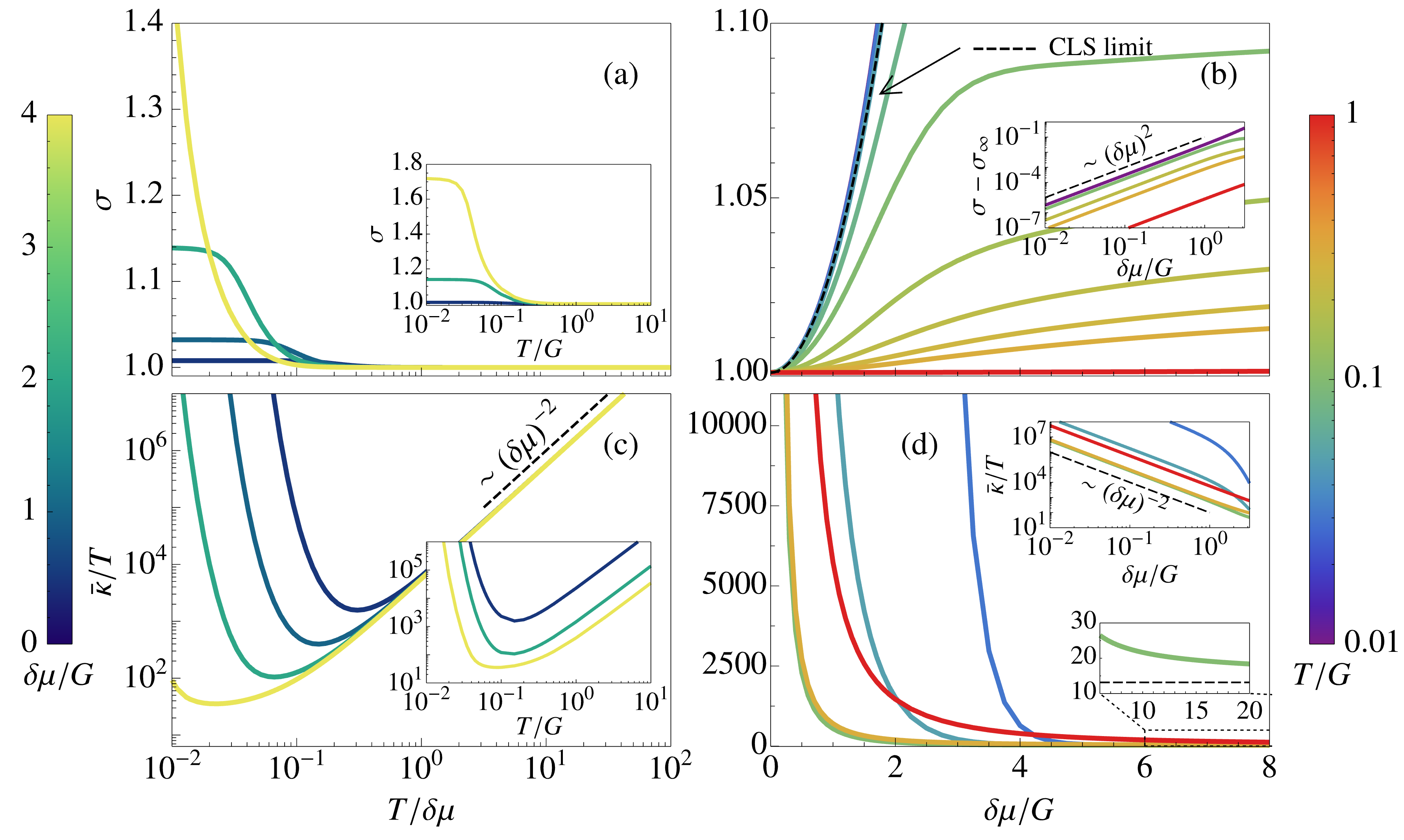}
    \caption{DC quantum critical transport in a 1D charge-neutral lattice, ${\mu(x) = \delta \mu \cos(G x)}$. Note that the thermoelectric conductivity ${\alpha = 0}$ (numerically confirmed, not shown) and thus $\bar{\kappa} = \kappa$. 
    \textbf{(a)} Electrical conductivity $\sigma$ as a function of $T/\delta\mu$ for various $\delta\mu/G$ (left colorbar). At low temperatures, %
    the lattice causes the conductivity to increase.
    Inset: electrical conductivity as a function of $T/G$.
    \textbf{(b)} Electrical conductivity $\sigma$ as a function of lattice amplitude for various temperatures $T/G$ (right colorbar). In the low-temperature limit for small $\delta \mu/G$, the scaling is that predicted by Chesler, Lucas and Sachdev \cite{chesler_conformal_2014}, (Eq. \cref{eq:sigma_CLS} black dashed line). Inset: difference from the CFT value $\sigma_{\infty}=1$ on a logarithmic scale, which grows as $(\delta \mu)^2$.
    \mbox{\textbf{(c)} Thermal} conductivity $\bar{\kappa}$ as a function of $T/\delta\mu$. There is a cross-over 
    from a universal high-$T$ regime to a strong thermal conductor-regime 
    at low $T$.
    Inset: thermal conductivity $\bar{\kappa}$ as a function of $T/G$.
    \textbf{(d)} Thermal conductivity $\bar{\kappa}$ as a function of lattice amplitude. The thermal conductivity decreases most rapidly with increased lattice amplitude around ${T/G \approx 0.1}$, which corresponds with the minima seen in panel (c). Upper inset: for weak lattices, the thermal conductivity scales as $1/(\delta \mu)^{2}$.
    Lower inset: for strong lattices, the thermal conductivity approaches the bound $\kappa/T \geq 4 \pi^2/3$ of \cite{grozdanov_incoherent_2016}. 
    }
    \label{fig:DC1Dtransport}
\end{figure}

There is one more expectation we can formulate.
Our system is on average isotropic in the thermodynamic limit. In quantum critical theories (with a holographic dual), this implies that the DC electrical conductivity is bounded from below as $\sigma_{DC} \geq \sigma_{\infty}$ \cite{grozdanov_absence_2015}. %
The electrical conductivity at charge neutrality in the absence of a lattice is %
given by that of the CFT dual to the Schwarzchild AdS$_4$ solution: $\sigma_{DC} = \sigma_{\infty}=1$. This saturates the bound. Thus, the only thing the conductivity can do is to increase with increased disorder/increased lattice amplitude around $\bar{\mu}=0$. We will indeed see this below.

\section{Thermoelectric transport in 1D charge-neutral lattices\label{sec:1Dresults}}

\subsection{DC transport in 1D charge-neutral lattices}
\label{sec:1D_DC_results}
To build on the results of Chesler, Lucas and Sachdev \cite{chesler_conformal_2014}, we first consider a 1D lattice at charge neutrality (i.e., that is charge-conjugation symmetric): we set the background chemical potential to be
\begin{equation}
    \mu(\Vec{x}) = \delta\mu \cos (G x)\, .
    \label{eq:chemical_variation}
\end{equation}
Recalling that the thermoelectric cross-conductivity $\alpha$ vanishes in a charge neutral setting, the remaining two DC transport coefficients, the electrical and thermal conductivity, $\sigma_{DC}$ and $\bar{\kappa}_{DC}$, are presented in \cref{fig:DC1Dtransport}. A first conclusion is that the presence of a lattice {\em increases} the conductivity, as should be expected based on the bound $\sigma_{DC}\geq \sigma_{\infty}$ from \cite{grozdanov_absence_2015}.
This effect is strongest at low temperatures for 1D lattices, as is highlighted \cref{fig:DC1Dtransport}(a). A similar temperature-behavior of the electrical conductivity $\sigma_{DC}$ was recently observed in a 
holographic model with fully random, strong disorder at finite density \cite{arean_disordered_2025}.
However, we wish to stress that the underlying physics of our model is completely different: at finite $\bar{\mu} \geq \delta \mu$, the DC transport is governed by the Drude peak --- leading to a greatly magnified overall scale --- whereas our increase for $\bar{\mu}=0$ stems from completely incoherent transport. Indeed, our observed scaling with respect to the disorder amplitude is inverted as compared with the results of Ref.~\cite{arean_disordered_2025}. Holographically, this difference is visible in that a finite $\bar{\mu} \neq 0$ leads to an AdS$_{2}\times\mathbb{R}^d$ near-horizon geometry, but that is not the case here.

\begin{figure}
    \centering
    \includegraphics[width=\linewidth]{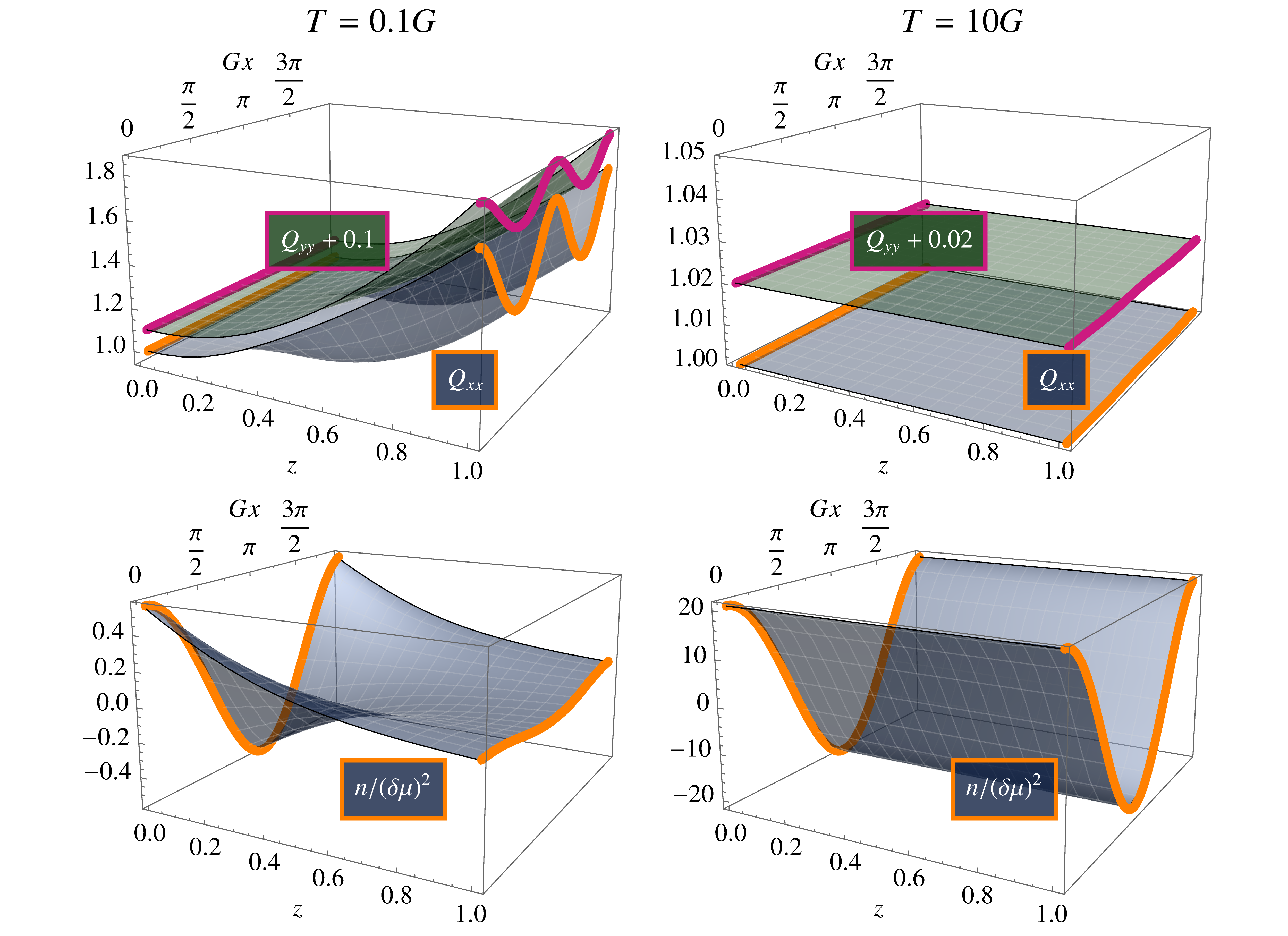}
    \caption{Metric components $Q_{xx}$ and $Q_{yy}$ (offset for clarity; top row) and local charge density $n(z) = F^{tz}$ (bottom row) for a 1D charge neutral lattice, $\mu(x) = \delta \mu \cos G x$ with $\delta\mu/G=2$. {The evolution in the $z$ coordinate captures the effect of RG flow.}
    At low temperatures $T=0.1G$ (left), the resulting $x/y$ anisotropy at the horizon gives rise to the asymptotic scaling in Eq. \cref{eq:sigma_CLS}.
    At high temperatures $T=10G$ (right), $\d s^2 \approx \d s_{\text{AdS}_4}^2$, and the 
    {renormalization of the charge density is cut off almost immediately.}
    The horizon charge density entering the Stokes equations, \cref{eq:stokes}, thus becomes ${n^H \approx \mu(x)}$ for $T\gg G$. 
    }
    \label{fig:field_components}
\end{figure}

This scaling of the electrical conductivity $\sigma_{DC}$ as a function of lattice strength was elucidated by Chesler, Lucas and Sachdev \cite{chesler_conformal_2014}: deforming the UV with a 1D lattice with $\bar{\mu}=0$ still results in an AdS$_4$ IR geometry at $T=0$,
except that in the natural UV coordinates the IR metric is anisotropic with regards to the $x/y$-directions. We highlight this effect in \cref{fig:field_components} (left column), that shows the RG flow (captured by the radial coordinate $z$), but halted due to the finite value of $T$, of the density vanishing in the IR and two %
metric components exhibiting the anisotropy.
Absorbing the emergent anisotropy in the coordinates to bring the IR metric into standard AdS$_4$ form also rescales the length scales of the dual theory, resulting in
\begin{equation}
    \sigma_{DC} = \sigma_\infty \lsb 1 + \frac{1}{2}\lp \frac{\delta\mu}{4 G}\rp^2\rsb\,.
    \label{eq:sigma_CLS}
\end{equation}
This perturbative result, valid for $T = 0$ and at $\delta \mu \ll G$, is shown as a black dashed line in \cref{fig:DC1Dtransport}(b) and which can be seen to asymptotically hold for $T \ll G$; in fact, the result extends to $\delta \mu/G = O(1)$ for low enough temperatures. 
At larger temperatures the deviation from $\sigma_{\infty}$ is not as strong, but still scales as $(\delta\mu)^2$ (inset, panel (b)).
Two comments about this strong coordinate anisotropy hiding translational symmetry mechanism are in order. Firstly, the explanation is particular to 1D, and will not apply to the 2D lattice as we shall see in \cref{sec:2Dresults}. Secondly, as explained by Chesler, Lucas and Sachdev, this emergent  translational symmetry is an artifact of the holographic large $N$ approximation \cite{chesler_conformal_2014}. 
There it was shown 
that by comparing to a system of $N_f$ free Dirac fermions in 2+1D at charge neutrality, the true ground state is not translationally invariant, but is a lattice of electron and hole filled regions. Tunneling transitions between these regions change the spectrum such that new Dirac nodes emerge \cite{parkNewGenerationMassless2008}. In the weak coupling description the enhancement of the DC electrical conductivity is due to the new low energy states in these new Dirac nodes. 

In the opposite limit where $T\gg G$, the high temperature cuts off any RG flow so early that the lattice %
barely modulates the metric at all, and the near-horizon geometry is essentially that of the isotropic AdS$_4$-Schwarzschild solution (\cref{fig:field_components}, right column).
The charge density, though strongly fluctuating on the horizon, is not able to
influence the bulk (spatially averaged over a lattical cell) electrical conductivity $\sigma_{DC}$: Setting $\d s^2 = \d s^2_\text{AdS$_4$}$, the conservation of the heat current
in the Stokes equations \cref{eq:stokes} reduces to a conservation of the velocity field,
\begin{equation}
    \nabla_i v^i = 0\, .
    \label{eq:diV=0}
\end{equation}
For 1D lattices where $i = x$, this enforces $v_x = \text{const.}$ and likewise, the continuity equation ${\nabla_i J^i = 0}$ together with the periodic boundary conditions constrain ${\nabla^i w = - n^H v^i}$. This amounts to a cancellation in the current \cref{eq:stokes_formulae} such that $J^i = E^i$ and the conductivity is trivially ${\sigma_{DC} = 1}$. As should be clear from \cref{eq:diV=0}, the fact that this equation has non-trivial solutions in 2D makes this story is completely different for the 2D lattices discussed in \cref{sec:2Dresults} below.

For the thermal conductivity $\bar{\kappa}_{DC}$, shown as a function of temperature in \cref{fig:DC1Dtransport}(c), we may conduct a similar analysis. At large temperatures --- roughly $T \gtrsim 0.3 G$ as can be deduced from the inset in \cref{fig:DC1Dtransport}(c) --- a hydrodynamical analysis ought to be valid, such that the thermal transport is fully coherent with a Drude-like response. This is what the numerical data indeed appears to reflect, with a decrease of thermal conductivity with increasing $T/G$ explainable by a thermal broadening of a Drude peak. However, a closer inspection of the AC thermal conductivity will reveal that this is not true.
Before we explain there the underlying physics, we do observe that at large $T\gg G$ the result is perturbative in the lattice strength $\delta\mu$: The thermal conductivity scales as $\bar{\kappa}_{DC}\sim (\delta\mu)^{-2}$ suggesting again (incorrectly) a single decay rate $\Gamma \sim (\delta\mu)^2$ (cf.~upper inset in \cref{fig:DC1Dtransport}(d)).
Increasing $\delta\mu$ therefore means decreasing $\bar{\kappa}$. However, it can be shown that the thermal conductivity is also bounded from below as $\kappa/T \geq 4 \pi^2/3$ \cite{grozdanov_incoherent_2016}, meaning that the thermal conductivity must eventually saturate. We indeed find that this is the case, as is shown in the lower inset in \cref{fig:DC1Dtransport}(d), where for the largest lattice strengths that we are able to probe ($\delta \mu= 20G$ at $T=0.1G$) the thermal conductivity stays above the bound by about a factor of $1.4$.

The thermal conductivity $\bar{\kappa}_{DC}$ attains a minimum around $T/G \simeq 0.1$ (inset, \cref{fig:DC1Dtransport}(c)), and as one lowers the temperature further, it enters a strong-thermal conductor regime where it increases exponentially and any semblance to (the $\omega\to 0$ limit of) the hydrodynamic Drude formula \cref{eq:hydro_transport} breaks down.
This point coincides with the onset of the increased electrical conductivity driven by the emergent strong horizon anisotropy as identified by Chesler, Lucas and Sachdev.
However, since this anisotropy is a coordinate artifact and the $T=0$ state of the system is dual to pure translationally invariant AdS$_4$, a perfect conductor is indeed expected to arise at the lowest temperatures.

\subsection{AC transport in 1D charge-neutral lattices}
\begin{figure}[t]
    \centering
    \includegraphics{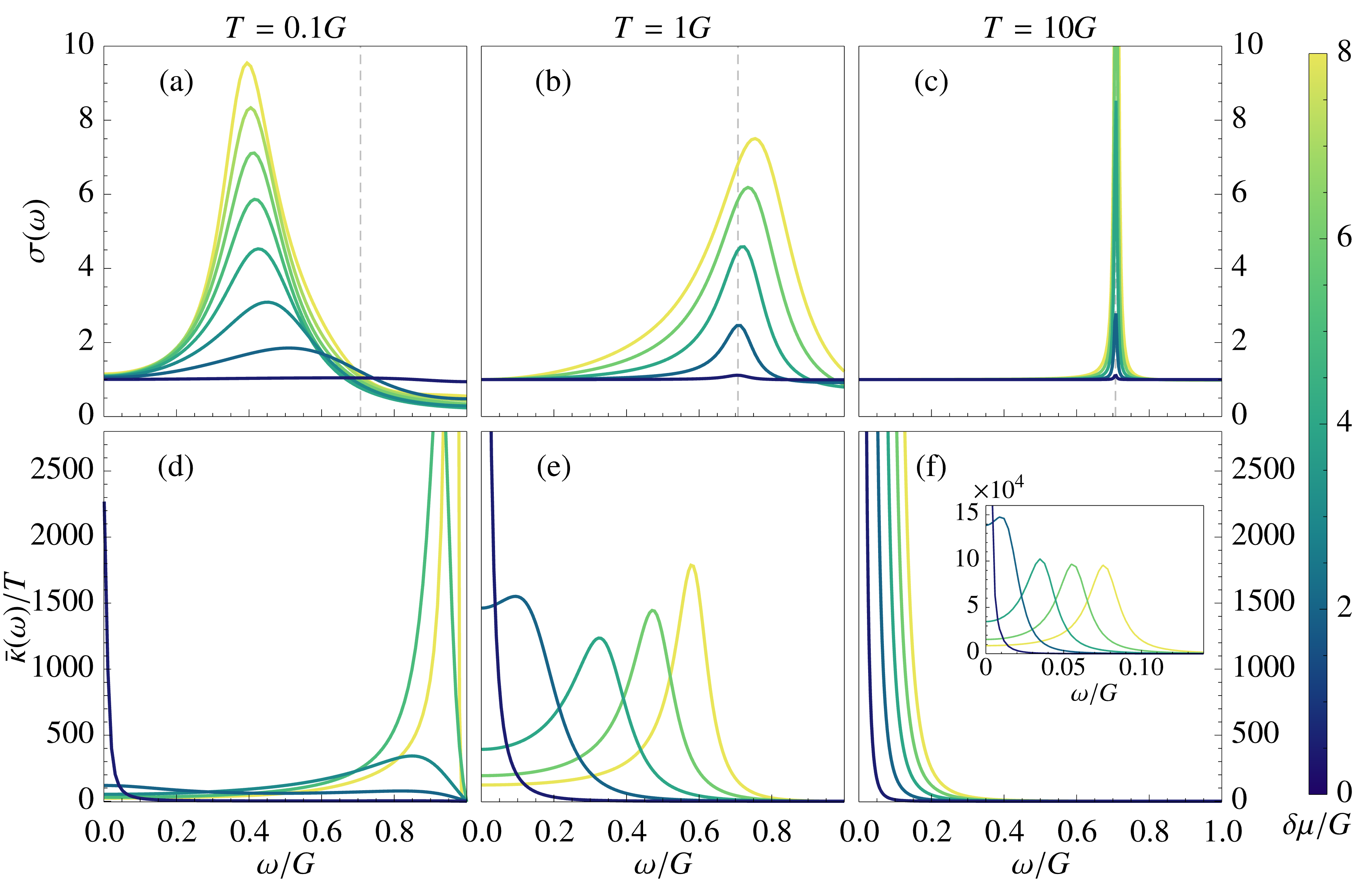}
    \caption{AC quantum critical transport in a 1D charge-neutral lattice, $\mu(x) = \delta \mu \cos (G x)$ at three different temperatures, ${T/G=0.1, 1, 10}$ (left to right). \textbf{(a-c)} The %
    AC electrical conductivity $\sigma(\omega)$ exhibits an Umklapp sound peak at $\omega \approx G/\sqrt{2}$ (gray dashed line), which is dominant at high temperatures. At lower temperatures (panel (a)), there is a slight downward shift in the peak frequency. \textbf{(d-f)} The AC thermal conductivity $\bar{\kappa}(\omega)$ is dominated by a pair of poles that moves toward $\omega = G$, a process which happens more rapidly at low $T/G$. Inset (f): Same data on a larger vertical scale for small frequencies. The peak moves closer to the origin, but is not a single Drude peak centered at $\omega=0$.}
    \label{fig:AC1DallT}
\end{figure}

Next, we study the finite frequency linear response.
In the high temperature regime, ${T\gg G}$, ${T\geq \delta \mu}$ shown in \cref{fig:AC1DallT}(b) and (c), the AC electrical conductivity $\sigma(\omega)$ is completely dominated by an Umklapped sound mode, which sits at $\omega =v_{\text{sound}}G \approx G/\sqrt{2}$ and which becomes increasingly sharp at larger $T/G$. The identification is readily made by comparing with systems at finite $\bar{\mu}$ \cite{donos_holographic_2014,balm_t-linear_2023}. This may seem surprising, as in a charge-neutral system one expects electrical and thermal transport to decouple, with the sound mode residing entirely in the thermal sector. This decoupling between electrical and thermal transport is however only valid at leading order in the charge density.
At subleading order (e.g., in terms of the density-density susceptibility $\chi_{nn}$), the cross-coupling between sound and charge diffusion does survive \cite{chagnet_hydrodynamics_2024}.

\begin{figure}[t]
    \centering
    \includegraphics{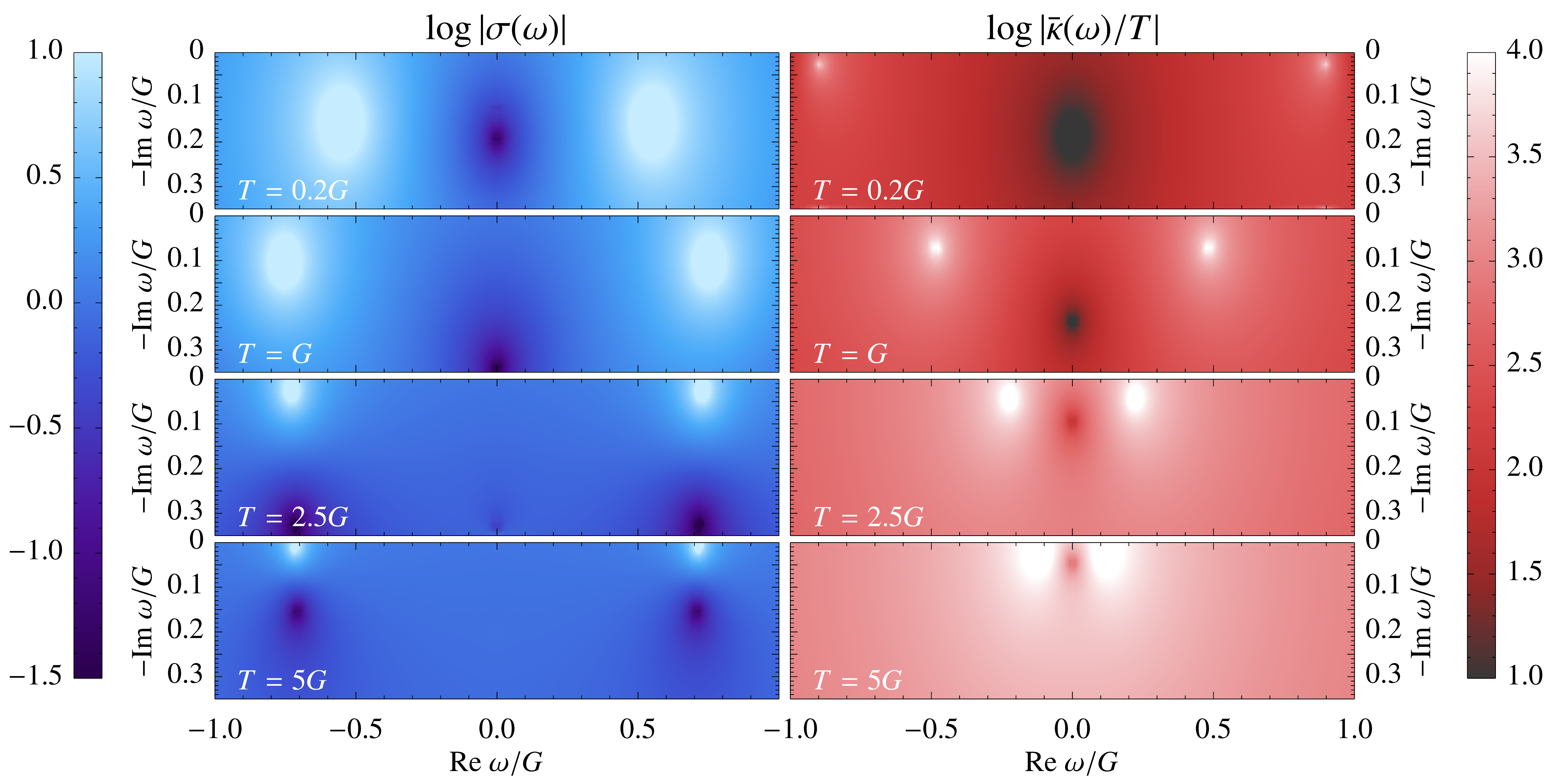}
    \caption{Pole structure of the AC response in a 1D charge-neutral lattice, $\mu(x) = \delta\mu \cos G x$ for various $T/G$. In all panels $\delta\mu = 6 G$. Left: logarithm of the complex electrical conductivity $\sigma(\omega)$. The Umklapp sound modes moves slightly toward the origin as $T/G$ is decreased for fix $\delta\mu/G$. There is no Drude pole or Umklapped charge diffusion pole present. Right: logarithm of the complex thermal conductivity $\bar{\kappa}(\omega)$. At all finite $\delta \mu$, transport is governed by a pair of modes that move in an arc toward $\omega = G$ as $T/G$ is decreased. This pair of modes arises from a collision of the Drude pole with the Umklapp charge diffusion pole. At charge neutrality Umklapp sound has negligible weight in the thermal channel and is indeed not visible.}
    \label{fig:AC1Dpoles}
\end{figure}

At these same temperatures $T\gg G$ the AC thermal conductivity $\bar{\kappa}(\omega)$ (panel (f)) seems at first to again be predominantly governed by a zero-frequency Drude peak, as expected from the hydrodynamic expression \cref{eq:hydro_transport}. 
As forewarned, however, this is not the case upon closer inspection. The zoomed-in inset at low frequencies in \cref{fig:AC1DallT}(f) shows that here are in fact two symmetric peaks at infinitesimally small $\omega=\pm \omega_{\text{peak}}$. To elucidate this more clearly we computed the response for complex-valued frequencies 
to find the exact location of the poles in the complex frequency plane. \Cref{fig:AC1Dpoles} shows indeed that there are two poles, each with a small but finite real part.

\begin{figure}
    \centering
    \includegraphics{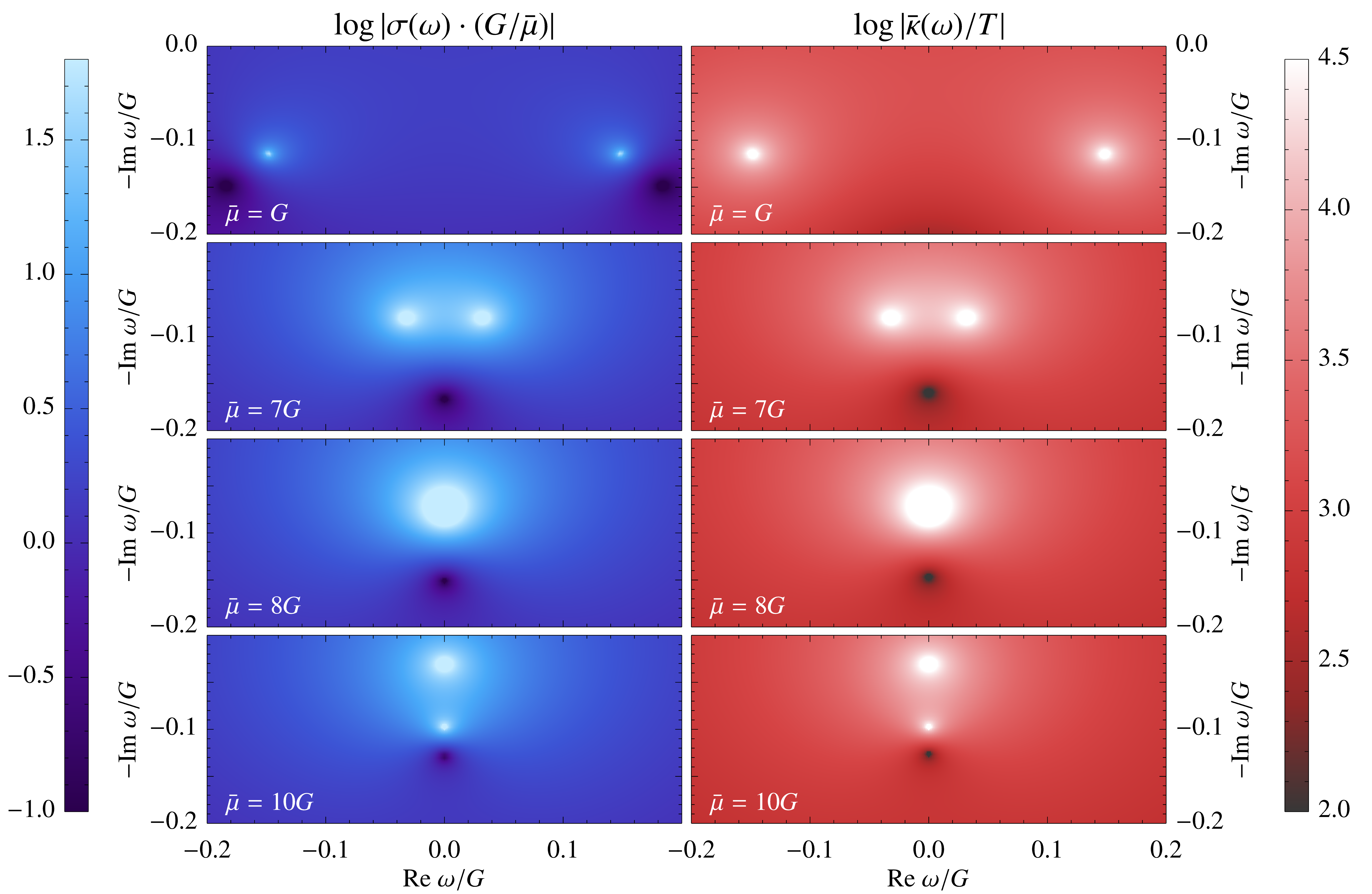}
    \caption{Pole structure of the AC response in a 1D lattice away from charge neutrality; ${\mu(x) = \bar{\mu} + \delta \mu \cos G x}$. Left/right: logarithm of the complex electrical/thermal conductivity for various $\bar{\mu}/G$, at temperature $T=G$ and with $\delta\mu=2 G$. At large enough finite $\bar{\mu}$, there exists both a Drude pole and a charge diffusion pole along the imaginary axis in both the electrical and thermal sector. These eventually collide, turning into the pair of modes that govern the thermal transport at $\bar{\mu}=0$. In the electrical sector, the residue of these poles vanishes as $\bar{\mu}\to 0$ due to a collision with a pair of zeros (top left panel).}
    \label{fig:poleCollision}
\end{figure}

These two poles must have arisen from a collision of two diffusive poles constrained to the imaginary frequency axis. The realization that Umklapp sound appears in the electrical conductivity $\sigma(\omega)$ points to an answer for the origin of these two poles. In the presence of spatial periodic modulations, hydrodynamic excitations are not given by Fourier modes, but by Bloch waves. For weak interactions these can be interpreted as Umklapped versions of original hydrodynamic excitations. In the case of charged hydrodynamics, this Umklapp effect predicts that the single Drude mode gets accompanied by two Umklapp sound modes and one Umklapp charge diffusion mode \cite{chagnet_hydrodynamics_2024}. Generically, both electrical and AC thermal transport will be dominated by these four modes at low frequencies. However, as first observed in \cite{balm_t-linear_2023}, for certain parameters the (diffusive) Drude mode and the Umklapp charge diffusion mode can collide and move symmetrically off the imaginary axis, acquiring a propagating part. For finite $\bar{\mu}$ this happens as $T/\bar{\mu}$ is increased. By adding a small overall chemical potential such that we're back in the regime where both modes are diffusive, we see in \cref{fig:poleCollision} that the same physics happens as we tune $\bar{\mu}/T$ to zero instead. 
This explains these two poles.
Umklapp hydrodynamics predicts four poles, but 
as $\bar{\mu}\rightarrow 0$ the residue of the two other poles --- Umklapp sound --- becomes negligible in the AC thermal conductivity $\bar{\kappa}(\omega)$ (cf.~\cref{fig:AC1Dpoles}). This can be confirmed analytically \cite{chagnet_hydrodynamics_2024}. Likewise, we see that at charge neutrality the residue of the Drude and charge diffusion pole in the AC electrical conductivity $\sigma(\omega)$ also vanishes. The set of four poles thus divide themselves between the electrical and thermal response at $\bar{\mu}=0$; Umklapp sound is purely in the electrical sector, whereas the diffusive poles (after the collision) govern the thermal response.

These exotic pole collisions near $\bar{\mu}/T=1$ are by now well known to occur in strongly coupled systems; for instance, it is also responsible for a vanishing plasma frequency for $\bar{\mu}/T \lesssim 1$ in models with dynamical electromagnetism \cite{gran_exotic_2019}.
For any finite $\delta \mu$, the thermal response is thus never truly Drude: unlike the hydrodynamic expressions \cref{eq:hydro_transport}, the momentum relaxation rate does not enter through a single Drude pole, but always through this pair.

This two-pole anchoring of the AC thermal conductivity $\bar{\kappa}(\omega)$ can be seen more clearly at intermediate temperatures $T=G$ (\cref{fig:AC1DallT}(e)).
The AC electrical response $\sigma(\omega)$ can still be seen to be dominated by Umklapped sound, but it has broadened significantly with respect to the high temperature regime. Qualitatively this broadening can be understood as an effect of stronger translational symmetry breaking as $\delta \mu/T$ has increased. The pinning of the Umklapp sound mode to the hydrodynamic prediction $\omega = v_{\text{sound}} G$ weakens as one approaches the viability-limit of the hydrodynamic approach $\omega \lesssim G$.

At even lower temperatures $T=0.1G$ (\cref{fig:AC1DallT}, panels (a) and (d)), the resonance peak in the electrical conductivity $\sigma(\omega)$ starts to move towards lower frequencies, 
becoming slightly more sharply peaked at higher $\delta \mu/G$ 
as compared to moderate temperatures.\footnote{Note that the increase in DC electrical conductivity with stronger lattices (\cref{fig:DC1Dtransport}) is barely visible in the $\omega \to 0$ limit in \cref{fig:AC1DallT} due to the scale.}
As is evident from \cref{fig:AC1Dpoles}, this response is still
dominated by the single pole that originates in Umklapp sound, though this regime is quite far from the Umklapp hydrodynamics regime $\delta\mu \ll T,\bar{\mu}$, and the imprint of the pole on the real axis is no longer close to $\omega =v_{\text{sound}}G$. 
This deformed Umklapp-sound peak remains absent in thermal transport: the  physical mechanisms between thermal and electrical transport stay distinct at low temperatures. 
In the AC thermal conductivity $\bar{\kappa}(\omega)$, on the other hand,  the mid-IR peak originating in the collision of the Drude pole with the Umklapp charge diffusion pole 
sharpens and moves towards $\omega = G$ as $T$ decreases. Though a peak at $\omega=G$ asks for an intuitive explanation in terms of the lattice and Umklapp, no such obvious explanation exists. Most such attempts will inadvertently rely on hydrodynamics, but for these lattice amplitudes one is outside its range of validity. A possible explanation is that this is a large $N$ artifact denoting a ``graviton-exchange'' in the holographic gravitational description as postulated in \cite{chesler_conformal_2014}. Such a collective mode would travel at the speed of light, and can therefore give rise to an $\omega=G$ excitation after Umklapp. %
As this happens at $T<G$, an $\omega =G$ energy-mode is outside of the hydrodynamical regime, precluding an easy understanding from the quantum critical theory perspective.

In conclusion, the a quantum critical model a charge neutrality in the presence of a 1D lattice exhibits two distinct regimes: at large $T/G$, we are in a 
regime with a fully incoherent DC electrical conductivity $\sigma_Q$, %
an AC electrical conductivity that is dominated by Umklapp scattering of the sound mode, and 
a thermal conductivity that at face value looks Drude-like, but is in fact dominated by a pair of modes slightly offset from zero frequency.
Conversely, at small $T/G$, an anisotropic horizon geometry results in a novel strongly coupled response regime characterized by two distinct mid-IR peaks independently examinable in the AC electrical and thermal conductivity, whereas both DC conductivities are %
explainable in the restoration of translational invariance in the true $T=0$ ground-state.

\section{Thermoelectric transport in 2D charge-neutral lattices\label{sec:2Dresults}}
We now turn to quantum critical transport in 2D lattices, encoded in a chemical potential on the form
\begin{equation}
    \mu(\Vec{x}) = \frac{\delta\mu}{2}\lsb \cos (G x)+ \cos (G y)\rsb\,.
    \label{eq:chemical_variation_2D}
\end{equation}
We will
show that there is a qualitatively novel contribution to the phenomenology of 
quantum critical transport in 2D charge-neutral lattices.
In addition, the electrical conductivity has a modified RG-scaling from 1D.
At the same time, the novel pole-collision physics 
behind 1D thermal transport does appear to continue to hold in 2D.
\subsection{DC transport in 2D charge-neutral lattices}
\begin{figure}[t]
    \centering
    \includegraphics{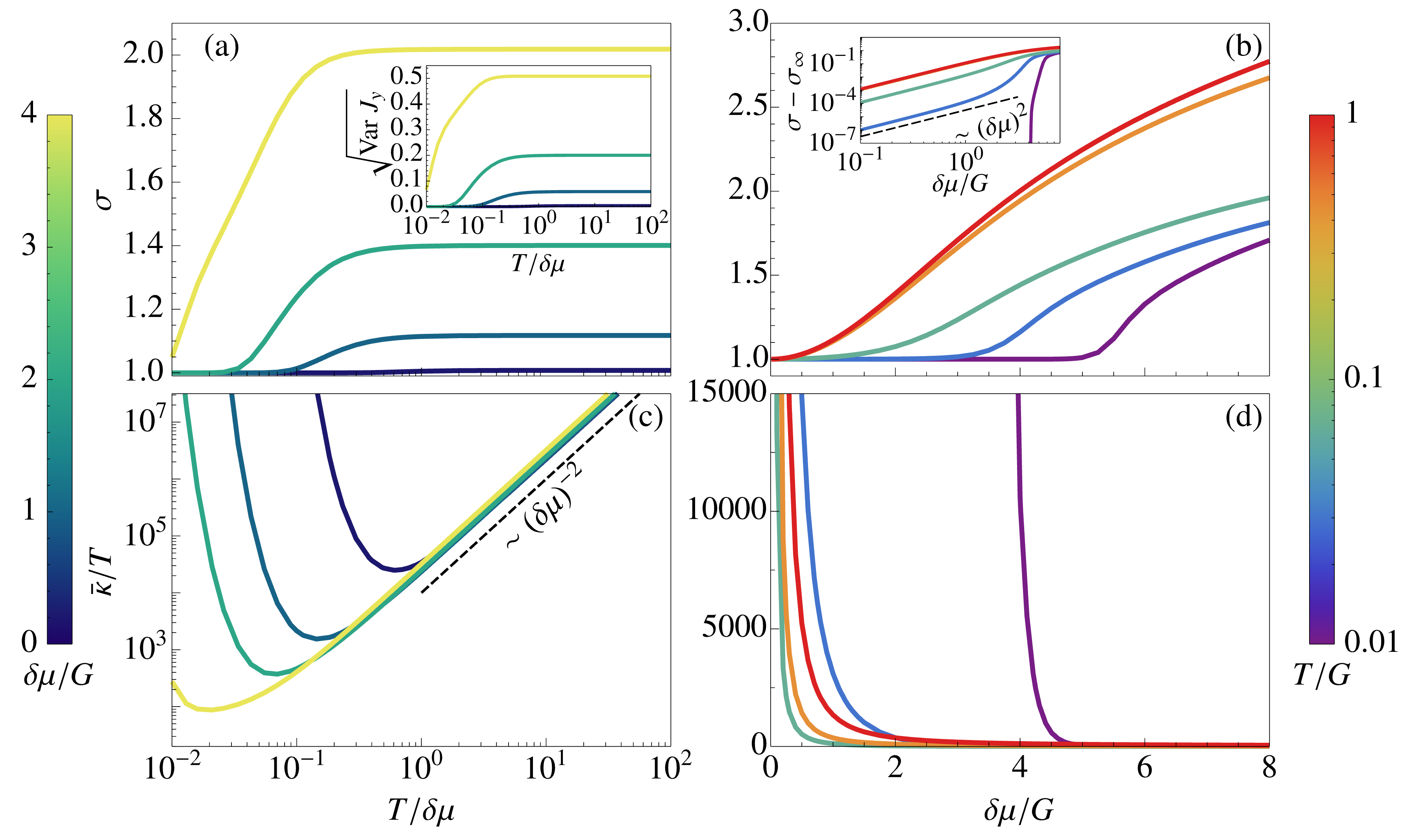}
    \caption{DC quantum critical transport in a 2D charge-neutral lattice, $\mu(\vec{x}) = \frac{1}{2}\delta \mu\left(\cos (G x) + \cos (G y)\right)$. Note that the thermoelectric conductivity ${\alpha = 0}$ (numerically confirmed, not shown). \textbf{(a)} Electrical conductivity $\sigma_{DC}$ as a function of $T/\delta\mu$ for various $\delta\mu/G$ (left colorbar). The lattice results in an increase in the conductivity at high temperatures.
    Inset: standard deviation of the fluctuations in $J_y(\vec{x})$ within a lattice cell, which correlates with the increased conductivity at high temperatures.
    \textbf{(b)} Electrical conductivity $\sigma_{DC}$ as a function of lattice amplitude for various temperatures $T/G$ (right colorbar). Inset: difference from the CFT value $\sigma_{\infty}=1$ on a logarithmic scale, which grows as $(\delta \mu)^2$.
    The thermal conductivity $\bar{\kappa}_{DC}$ as a function of temperature \textbf{(c)} and lattice strength \textbf{(d)} remains similar as for 1D lattices.}
    \label{fig:DC2Dtransport}
\end{figure}
The difference 
in electrical transport is immediately evident in the DC conductivity $\sigma_{DC}$, as shown in \cref{fig:DC2Dtransport}
(panel (a)): 
the electrical conductivity now increases with temperature
instead of decreasing monotonically back down to the translationally invariant quantum critical value $\sigma_{\infty}$.
This remarkable enhancement of the conductivity at higher temperatures can be understood if the lattice modulation acts as an irrelevant deformation. For a spatially dependent operator this is not straightforward to determine, but there is heuristic argument that suggests that this is indeed so.
One can make an analogy between a periodic source of translational symmetry breaking which vanishes averaged over a lattice cell 
$\int_{\vec{x}}^{\vec{x}+\vec{a}} \mu(\vec{x})=0$
but whose square does not 
$\int_{\vec{x}}^{\vec{x}+\vec{a}} \mu(\vec{x})\mu(\vec{x})\neq 0$, 
and random disorder with 
$\langle g (\vec{x})\rangle=0$, 
$\langle g (\vec{x}) g(\vec{x}')\rangle=g^2\delta(\vec{x}-\vec{x}')$. 
For such generic Gaussian random field disorder introduced through $\int\!\text{d}\vec{x}\text{d}t\, g(\vec{x}){\cal O}(\vec{x})$, the Harris criterion \cite{harris_effect_1974} reveals
that it is relevant if the scaling dimension of ${\cal O}$ is ${\Delta_{\cal O} < \frac{1}{2}(d+1)}$. 
For a randomly disordered chemical potential modulation around a charge neutral lattice coupling to ${\cal O}(\vec{x})=n(\vec{x})$ with dimension $[n]=d$, the modulation saturates this bound for $d=1$, but violates it for $d \geq 2$. Such %
chemical potential disorder
is thus marginal in $d=1$ and irrelevant in higher dimensions. 
Although we here consider lattice modulations around zero average rather than random disorder,
our numerics support this same division, suggesting that the vanishing of averages makes the lattice modulation act like disorder and that this is a good hint towards the underlying physics.\footnote{As can be seen in \cref{fig:field_components}, the lattice modulation is marginally \emph{irrelevant} for 1D lattices. The charge-density profile exhibits the same qualitative behavior for 2D lattices.
}

This is further supported by an analysis of the near-horizon geometry that determines the DC response.
Similar to 1D lattices, for $T=0$ the metric remains purely AdS$_4$ in the IR. However, for the square lattice we use, this is an isotropic AdS$_4$ and there is no enhancement expected of the quantum critical conductivity. Indeed our numerics show that the 2D DC electrical conductivity approaches $\sigma_{\infty}$ as temperature decreases. This must be from a higher value $\sigma_{T}>\sigma_{\infty}$ by the bound \cite{grozdanov_absence_2015}, thus following an irrelevant RG flow. We do note
that we 
are numerically limited by our numerics to go to temperatures lower than $T/\delta\mu = 0.01$ while keeping $\delta\mu/G$ reasonably large.

At the same time, the arguments for a trivial conductivity $\sigma_{DC} = \sigma_{\infty}$ {\em at high temperatures} in the case of 1D must break down for 2D lattices, as is readily confirmed. In particular, from the holographic point of view, \cref{eq:diV=0} now allows for spatial variations in the horizon velocity field.
The Stokes equations \cref{eq:stokes} reduce in this limit to
\begin{align}
    v^i \partial_i \rho^H + \partial_i (E^i + \partial^i w) &= 0 \,,\\
    \partial^2 v_i + 4 \pi T \zeta_i - \partial_i p + \rho^H(E^i + \partial^i w) & = 0\,,
\end{align}
where we have used that all transport coefficients but $\rho^H$ become trivial and that $\partial_i v^i = 0$.
Thus, while the unit cell average of currents perpendicular to the electric field $E_x$ must vanish ($\bar{J}_y = 0$), 
local variations of $J_y(\vec{x})$ need not be zero.\footnote{Recall that $\bar{\mu}=0$ causes the global Hall conductivity ($\sigma_{xy}=0$) to vanish, but local variations are still allowed.}
In fact, at high temperatures it will in general fluctuate, whereas at low temperatures we find $J_y(\vec{x}) \to 0$. We show in the inset in \cref{fig:DC2Dtransport}(a) that the variance of $J_y(\vec{x})$ indeed follows the same general shape as the electrical conductivity. 
We may interpret this variance in $J_y(\vec{x})$ as the consequence of the current following
a curved  
path through the potential landscape, in such a way that the conductivity is increased. Though a holographic computation in the strong coupling regime cannot reveal this, the natural mechanism prompting such perpendicular variations in a weakly coupled (Boltzmann transport) system that grow with temperature would be thermally assisted fluctuations or thermally assisted local scattering. 

\begin{figure}
    \centering
    \includegraphics[width=\linewidth]{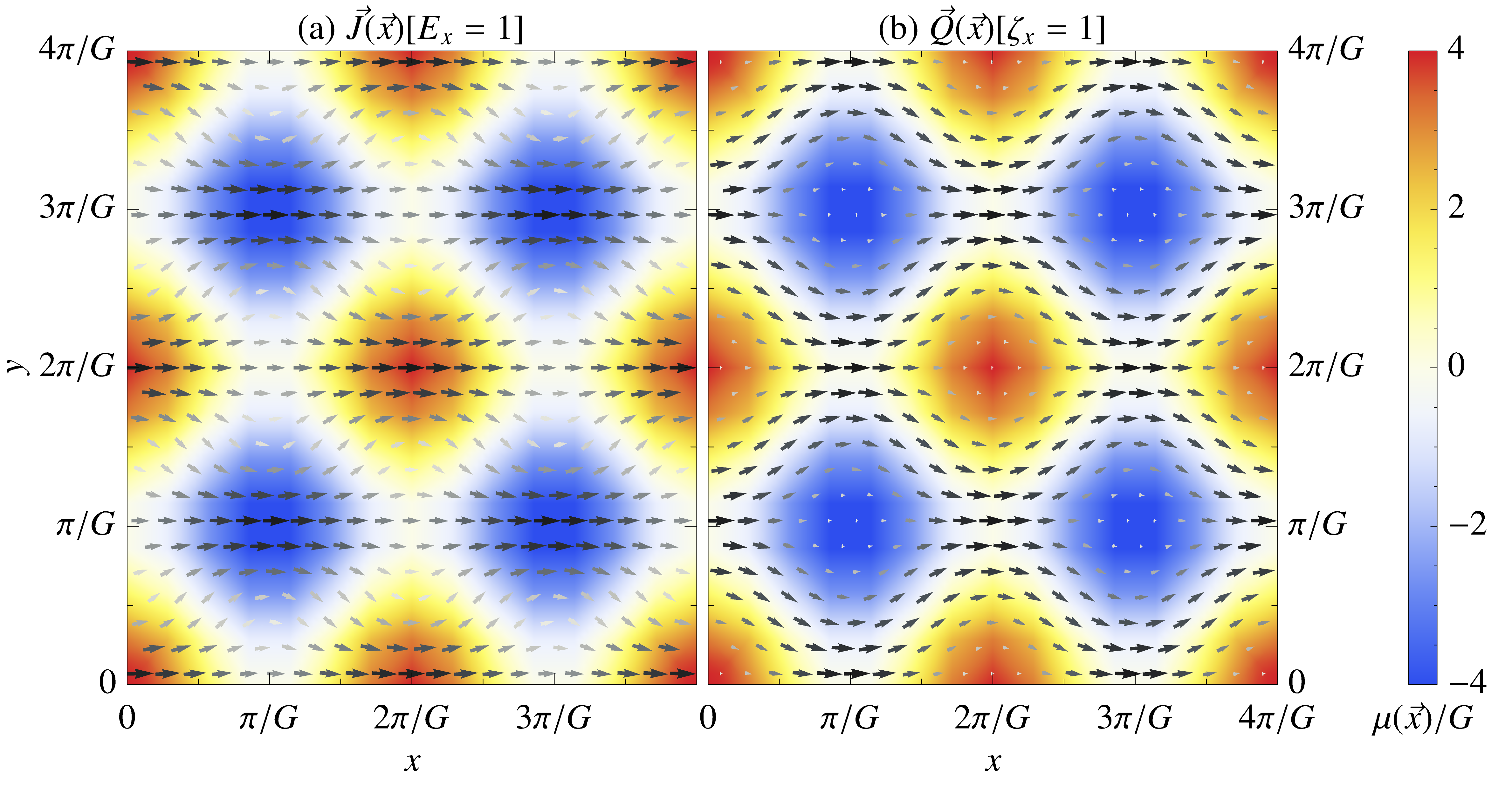}
    \caption{Vector plot of the electrical and thermal currents on the horizon for a 2D lattice with $\delta \mu/G = 4$ at a moderately high temperature $T=G$. Darker, larger vectors indicate greater magnitude. The underlying heat map shows the varying chemical potential, which is proportional to the horizon charge density.
    \mbox{\textbf{(a)} Local} electric current $\vec{J}(\vec{x})$ when applying a constant electric field in the $x$-direction, capturing the behavior of the electrical conductivity $\sigma$. The current flows from regions of $\mu(\vec{x})\approx 0$ to where $\abs{\mu(\vec{x})}$ is large. \textbf{(b)} Local thermal current $\vec{Q}(\vec{x})$ when applying a constant thermal gradient in the $x$-direction, capturing the behavior of the thermal conductivity $\bar{\kappa}$. The thermal conductivity is largest when $\mu(\vec{x})\approx 0$.
    }
    \label{fig:stokesHorizon}
\end{figure}
This picture is corroborated by a direct inspection of the current flows.
\Cref{fig:stokesHorizon} shows both the unrenormalized lattice as well as the local electrical and thermal currents $J_i(\vec{x})$ and $Q_i(\vec{x})$ on the horizon, representing the quantities solved for in the Stokes equations \cref{eq:stokes}.\footnote{Note that the extraction of the physical currents in the dual quantum critical theory from the horizon currents is based on the fact that the unit cell averages $\bar{J}_i,\bar{Q}_i$ do not renormalize. The horizon values of the local expressions $J_i(\vec{x})$ and $Q_i(\vec{x})$ do renormalize and are in general not identical to their boundary counterparts. Nevertheless, the Stokes equations describing the problem give insight into the underlying physics. We do in fact find a similar pattern in the boundary currents $J_i(\vec{x})$ and $Q_i(\vec{x})$ as computed by a (more numerically complicated) AC calculation for $\omega\to 0$.}
From this, an interesting picture emerges, which on physical grounds must have its origins in that the collision-less to hydrodynamic crossover in quantum critical systems can happen at larger scales than one would expect: it seems that each patch in the transverse plane can be thought of as a local (electric) Reissner-Nordstr{\"o}m geometry, i.e., locally it is a collective quantum critical system at finite density, even though the global charge density vanishes. If so, one can consider the hydrodynamic results for the conductivities \cref{eq:hydro_transport}, assuming the expressions to be valid at some (suitably defined) local point $\vec{x}$, which should hold in an adiabatic approximation where the spatial variation is on a much larger scale than the onset of hydrodynamics:\footnote{
As the system is in thermal equilibrium, the temperature $T$ will still be $\vec{x}$-independent.}
\begin{align}
     \sigma(\omega, \vec{x}) &= \frac{n^2(\vec{x})}{\varepsilon(\vec{x}) + p(\vec{x})} \frac{1}{\Gamma(\vec{x}) - i \omega} + \sigma_Q(\vec{x}) \,, \\
     \bar{\kappa}(\omega,\vec{x}) &= \frac{s^2(\vec{x}) T }{\varepsilon(\vec{x}) + p(\vec{x})} \frac{1}{\Gamma(\vec{x}) - i \omega} + \frac{\mu^2(\vec{x})}{T}\sigma_Q(\vec{x})\, .
    \label{eq:hydro_transport_local}
\end{align}
Since for a finite density system, the electrical Drude weight 
is given by the (local) density squared, 
the local conductivity $\sigma_{DC}(\vec{x})$ is insensitive to the sign changes of the local chemical potential. This can be seen in \cref{fig:stokesHorizon}(a): the electrical current density $J_i(\vec{x})$ concentrates in regions of large $\abs{\mu(\vec{x})}$ independent of the sign, and evades 
regions where $\mu(\vec{x}) \approx 0$ (where the local conductivity is $\sigma = \sigma_\infty = 1$). This is reminiscent of Effective Medium Theory, where the current takes the path of least resistance through a material with regions of locally strongly varying conductivity \cite{stroud_generalized_1975}.\footnote{Note that this means that locally, transport is governed by a Drude-like pole, but there is no global hydrodynamics where a Drude mechanism can explain transport.}

This same evading flow patterns are present in local probes of the thermal horizon current, albeit with an important twist. 
\Cref{fig:stokesHorizon}(b) shows the thermal current $Q_i(\vec{x})$ when applying a constant thermal gradient in the $x$-direction. 
Keeping with our picture of local Reissner-Nordstr{\"o}m geometries and {\em local} hydrodynamic transport, 
the thermal conductivity $\bar{\kappa}(\vec{x})$ will be proportional to $\frac{s(\vec{x})^2T}{{s(\vec{x}) T}+{\mu(\vec{x}) n(\vec{x})}}$ upon approximating $\epsilon+P=sT+ \mu n $, cf.~\cref{eq:hydro_transport_local}.
Similar to the electrical conductivity, 
the thermal conductivity is therefore insensitive to the local sign of the chemical potential: 
however, now regions of large $\mu(\vec{x})$ are regions of weak transport and vice versa. 
Hence one has a weaker thermal current at large local $\abs{\mu(\vec{x})}$, and a stronger current in regions of small local $\abs{\mu(\vec{x})}$, in contrast to the electrical conductivity.\footnote{The thermoelectric conductivity $\alpha$ --- not shown here for brevity --- is on the other hand sensitive to the sign of the local chemical potential and density (cf.~\cref{eq:hydro_transport}) and indeed, we find that the local current $J_i(\vec{x})$ when applying a thermal gradient (or thermal current $Q_i(\vec{x})$ for an external electric field) flows in a direction set by $\text{sgn}(\mu(\vec{x}))$, leading to $\alpha=\bar{\alpha} = 0$ upon averaging (since $\bar{\mu}=0$).}

Despite this far more intricate picture of the 2D flow patterns at sub-lattice length, the effective overall averaged DC thermal conductivity $\bar{\kappa}_{DC}$, shown in \cref{fig:DC2Dtransport}(c), exhibits the same phenomenology of the 1D periodic quantum critical DC thermal conductivity (\cref{fig:DC1Dtransport}(c)). At large $T/G$, the scaling is yet again 
perturbative in the lattice amplitude suggestive of a momentum relaxation rate ${\Gamma \sim \delta\mu^2}$,
transitioning to a strong thermal conductor at low $T/G$ due to the irrelevancy of the lattice. As we shall see shortly, this suggestion is again deceptive, as the AC thermal conductivity $\bar{\kappa}(\omega)$ will reveal that also here multiple poles contribute.

As already pointed out, the 2D electrical conductivity shows a very different behavior with a thermally activated quantum critical conductivity enhancement at larger $T/G$ flowing to the purely homogeneous charge neutral quantum critical $\sigma_{\infty}$ conductivity at low $T/G$ (\cref{fig:DC2Dtransport}(a)).

\subsection{AC transport in 2D charge-neutral lattices}
\begin{figure}[t]
    \centering
    \includegraphics[width=\linewidth]{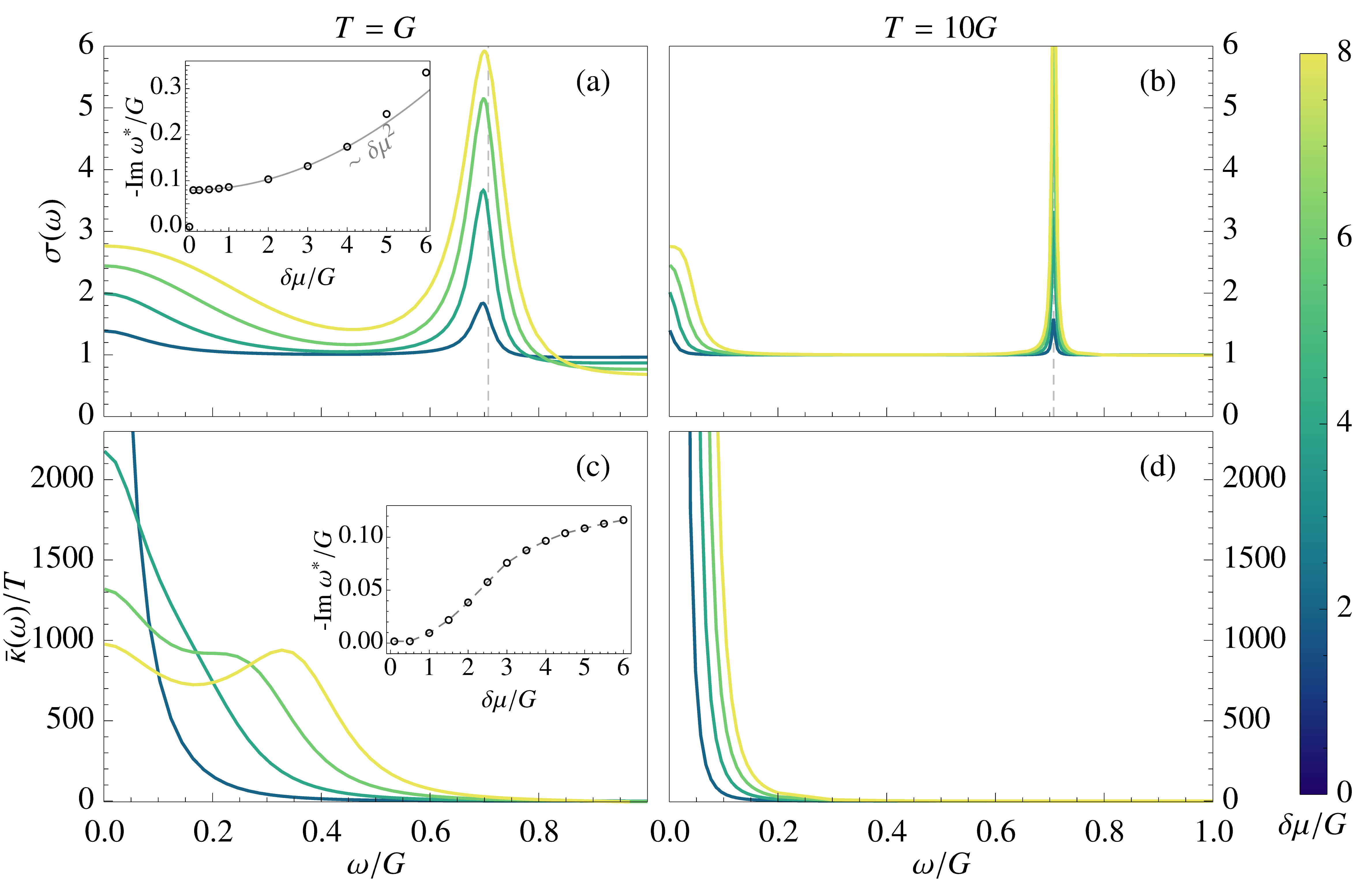}
    \caption{AC transport in a 2D charge neutral lattice, $\mu(\vec{x}) = \frac{1}{2}\delta \mu\left(\cos (G x) + \cos (G y)\right)$. \textbf{(a-b)} In contrast to the 1D lattice, the 2D AC electrical conductivity $\sigma(\omega)$  exhibits a weak zero-frequency peak at large lattices, in addition to the the Umklapp sound peak at $\omega \approx G/\sqrt{2}$ (gray dashed line).
    Inset: location of the new zero-frequency diffusive pole $\omega^*_{\text{elec}}$ along the imaginary axis. The solid gray line is a fit on the form $-\Im \omega^* = a + b \,\delta\mu^2$ over the interval $\delta\mu/G \in (0,4]$. \textbf{(c-d)} As for 1D lattices, the 2D AC thermal conductivity $\bar{\kappa}(\omega)$ shows a similar mid-IR peak at larger lattice strengths arising from the collision of the Drude pole with the Umklapp charge diffusion pole. There is also an additional novel zero-frequency peak as for the electrical conductivity. Inset: location of the new zero-frequency diffusive pole $\omega^*_{\text{therm}}$ along the imaginary axis. The gray dashed line is a guide to the eye. The scaling with $\delta \mu$ is different from the corresponding pole in the electrical conductivity, and is thus of a different origin.
    }
    \label{fig:AC2dhighT}
\end{figure}
For the AC transport in a 2D lattice, we are forced to restrict our focus to temperatures $T\gtrsim G$ due to numerical limitations.
The AC response at $T=G$ (at which point the asymptotic high-temperature behavior in the DC response has set in, cf.~\cref{fig:DC2Dtransport}) is shown in \cref{fig:AC2dhighT}. By comparing with the 1D results (\cref{fig:AC1DallT}), the AC results show that this is nevertheless an intermediate temperature regime. 

In the AC transport 
we encounter a qualitatively new contribution in 2D compared to 1D. \Cref{fig:AC2dhighT}(a) shows that the AC electrical conductivity now displays classic bad metal behavior at low frequencies up the Umklapp sound peak.
It therefore appears that a new zero-frequency diffusive mode mixes in, which may be
identified by studying the response at purely imaginary frequencies. 
In the inset in \cref{fig:AC2dhighT}(a) we show the location of this pole along the imaginary frequency axis. 
For finite $\delta\mu$ it is well described by a quadratic function consistent with a perturbative origin in the lattice, aside from a finite offset at $\delta \mu=0$. Such a discrete jump at $\delta \mu =0$ is most likely attributable to a mode collision for tiny $\delta\mu \ll G,T$. Within our numerical resolution we could not resolve or identify such another mode within the electrical conductivity, but they are there as we shall see.\footnote{
An added subtlety is that the peak has a Fano lineshape. As can be seen in the 1D lattice pole landscape, \cref{fig:AC1Dpoles}, there also exists a zero of the electrical conductivity along the imaginary axis. The same is true in 2D, and at small enough $\delta\mu \ll G,T$ the pole and the zero appear to collide.}
The bad metallic behavior persists at higher temperatures (\cref{fig:AC2dhighT}(b)). As $T$ increases, the peak height does not change, consistent with the results for DC transport in \cref{fig:DC2Dtransport}(a), where the DC value can be seen to attain a constant value for $T \gtrsim G$. The peak gets increasingly sharp, however, in contradistinction to thermal broadening, illustrating its unconventional origin.

A new weak zero-frequency peak is also visible in the 2D AC thermal transport (\cref{fig:AC2dhighT}(c)) on top of the mid-IR mode arising from the collision of the Drude pole with the Umklapp charge diffusion pole.
A first impulse might be to surmise that this novel mode is the same novel Drude-like mode as in the electrical conductivity,
suggesting that some mixing between thermal and electrical conductivity is starting to occur. This is unlikely, however, as the AC thermoelectric conductivity $\alpha_{xx}(\omega)$ remains strictly zero  within our numerical precision,
indicating that the transport mechanisms of thermal and electrical transport remain distinct. 
The new pole in the AC thermal conductivity can therefore not be the same new pole as in AC electrical transport. It is indeed a different pole, as can be seen in the inset in \cref{fig:AC2dhighT}(c): it does not exhibit the same scaling with $\delta\mu$ as the similar pole in the electrical sector. %
At larger temperatures, $T\gg G$ (\cref{fig:AC2dhighT} (panel (d)), its contribution to the DC thermal conductivity is subleading to that of the collided Drude and Umklapp charge diffusion poles, whose real part starts to get closer and closer to $\omega=0$, showing a similar high-$T$ scaling of the 1D and 2D DC thermal conductivities.

With our understanding of the DC conductivities through an Effective Medium flow through channels of least resistance, we can attempt a qualitative explanation of the underlying physics. The essence of this Effective Medium type point of view is that one essentially has a composite of independent materials --- in this case particle-filled and hole-filled regions created by the periodic chemical potential. Each independent region has a single fluid hydrodynamic description, meaning that there are effectively as many Drude modes as there are local fluid regions that weakly interact with each other. This viewpoint is supported by similar findings in charge-neutral graphene, where in a periodic background one has new Dirac points that are remnants of the local Fermi surface in each pocket \cite{parkNewGenerationMassless2008,chesler_conformal_2014}. In a collective interacting system, the same should happen for the emergent hydrodynamics provided the lattice period is large enough and the boundaries sharp enough. In a 1D charge neutral periodic lattice these modes act in unison, however; most of the local contributions either add coherently or cancel each other perfectly and a single global hydrodynamic description survives. In particular, there is no net charge flow leaving only a quantum critical electrical conductivity. In 2D, however, thanks to variance of the flow allowed by the perpendicular direction, this cancellation of each of the local hydrodynamic charge flow is no longer exact and a remnant Drude-like mode $\omega^\ast_{\text{elec}}$ survives. This mode exists in addition to the global Drude mode, which happens to collide with the global Umklapp charge diffusion mode. This ``existence of more modes than conserved quantities'' would be the hydrodynamic counterpart of the violation of the RG $c$-theorem in charge-neutral graphene in a periodic background. Naively, $c$ counts the number of massless Dirac modes, which in charge-neutral graphene in a periodic potential increases due to the emergence of the new Dirac points. This violation is allowed because the $c$-theorem relies on translational invariance, which is no longer present.

Similarly, in the thermal sector a remnant Drude-like mode $\omega^\ast_{\text{therm}}$ survives from the mesoscopic physics. As we have seen this is a different mode than the remnant Drude-like mode in the electrical sector. Since all such novel Effective Medium type modes should interact, because they exist in region of non-zero local $\mu(x)$,
simplicity argues that it is the level repulsion between this mode and the remnant Drude mode $\omega^\ast_{\text{elec}}$ that causes the offset of the latter at $\delta \mu=0$. At the same time, it is the vanishing of the residue of each respectively that causes only one of them to arise in the respective transport currents.

On top of this novel emergent Drude-like mode of mesoscopic Effective Medium origin, we do recognize the same characteristics underlying 2D charge neutral quantum critical transport as in 1D. In particular, in the AC electrical conductivity $\sigma(\omega)$ at intermediate frequencies, we still see the Umklapped sound mode at $\omega = G/\sqrt{2}$.

\subsection{DC magnetotransport in 2D charge-neutral lattices}
\begin{figure}[t]
    \centering
    \includegraphics{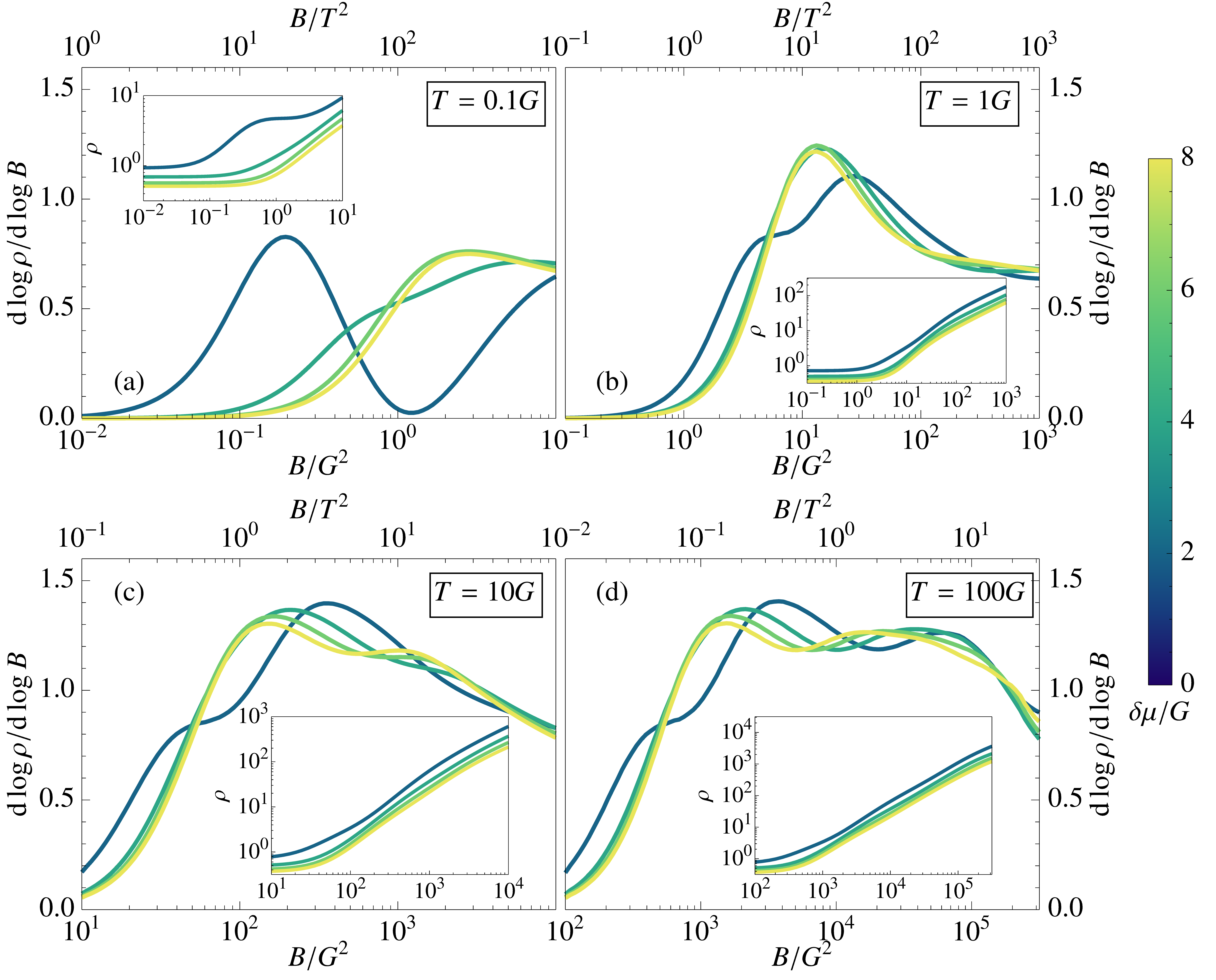}
    \caption{Logarithmic derivative of the longitudinal magnetoresistance $\rho$ with respect to the magnetic field, capturing the local scaling exponent $\rho \sim B^\beta$. Insets: longitudinal magnetoresistance as a function of the magnetic field $B$.}
    \label{fig:magneto}
\end{figure}
A novel aspect of 2D transport is that we can now also consider transport in the presence of a background perpendicular magnetic field $B_z$. We limit our discussion to DC transport only, due to the complexity of the linearized equations of motion for the 
time-dependent
perturbations in the case of $B\neq 0$.\footnote{A text file containing the simplified equations of motion and horizon boundary conditions for the perturbations totals about 40 MB.}

Holding all other parameters ($T/G, \delta \mu/G)$ fixed, the most striking result is an almost linear dependence of the longitudinal magneto-resistance $\rho_{xx}=\frac{1}{\sigma_{xx}}$ on the magnetic field $B/T^2$ beyond $B\sim T^2$ independent of the size of $\delta \mu/G$: 
in \cref{fig:magneto}, we plot the logarithmic derivative of $\rho(B)$ to extract the local scaling exponent $\beta$ for $\rho_{xx} \sim \abs{B}^\beta$. In the ``Effective Medium regime'', $T\gtrsim G$, we obtain roughly linear dependence on the magnetic field. As we increase the temperature further, we find that the scaling 
stays quasi-linear over a larger regime.
Recalling that parity implies that the longitudinal resistivity can only be a function of $B^2$, the explanation might therefore be a quadrature-type relation ${\rho_{xx}\sim ({\rho_{xx}(B=0)^{2/\beta}+\gamma \mu_B B^2})^{\beta/2}}$ well-known to occur in Effective Medium Theory \cite{parish_non-saturating_2003,hayes_scaling_2016,ramakrishnan_equivalence_2017,boyd_single-parameter_2019,ayres_incoherent_2021}. Notably, we obtain emergent $B$-linear scaling without an ad hoc addition of mesoscopic disorder on top of our microscopic model. As we argued in the previous sector, the chemical potential modulation in a charge neutral background suffices to cause a flow to a ground state that consists of effectively independent particle-filled and hole-filled regions.

\begin{figure}[t]
    \centering
    \includegraphics{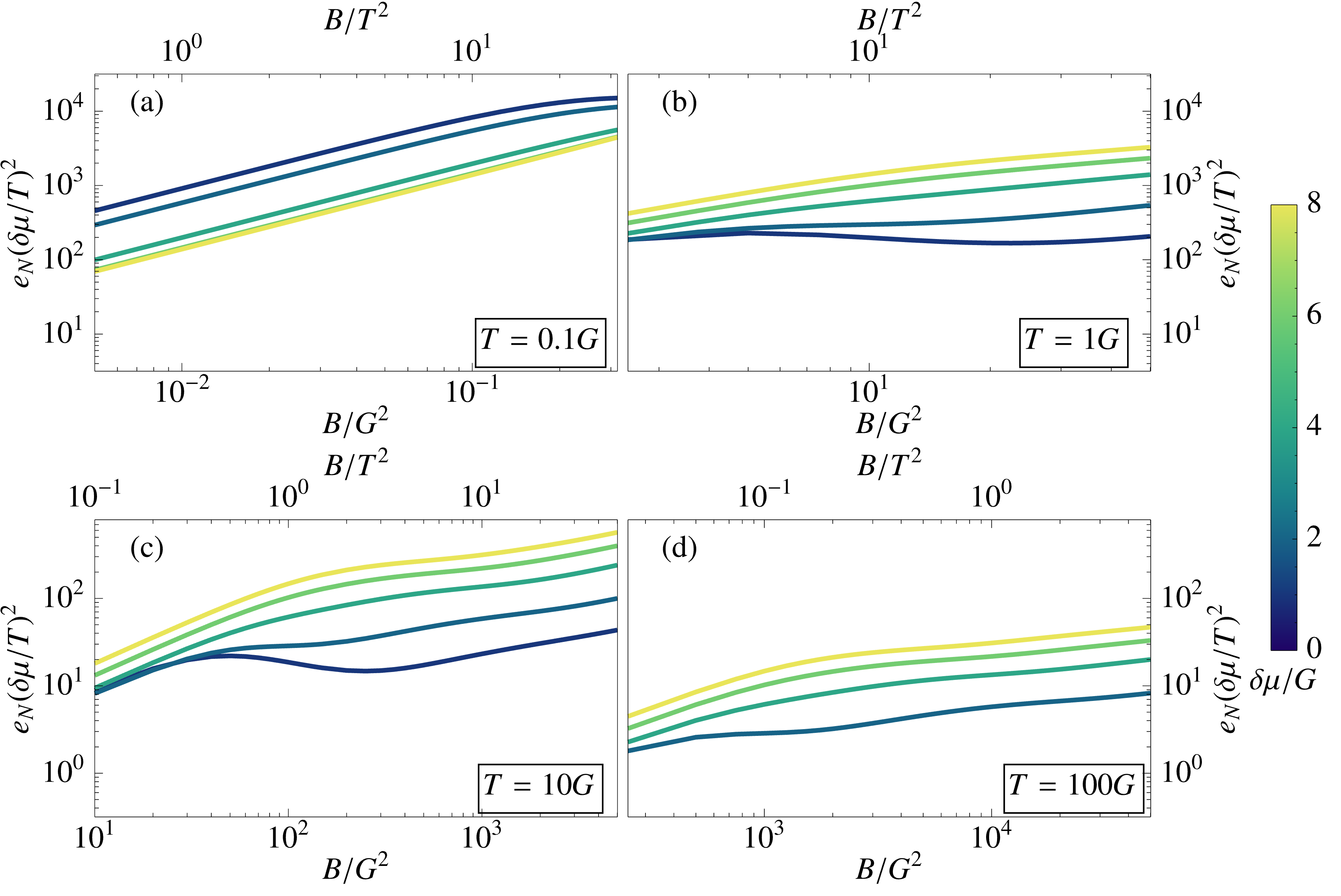}
    \caption{Nernst coefficient $e_N$ at 4 different temperatures; $T=(0.1,1,10,100)G$ \textbf{(a-d)}.}
    \label{fig:nernst}
\end{figure}

Although there is no Hall effect in a 2D charge-neutral lattice, the parity-breaking by the magnetic field can induce a Nernst effect. As recalled in \cref{eq:Nernst_hydro}, for a charge neutral system in single-fluid hydrodynamic regime at large $T/\delta \mu$ dominated by a single Drude mode, this measures the momentum relaxation rate directly as $e_N=B/T\Gamma$. Though our results have shown that this is not applicable here, we have shown that in this regime, transport is perturbative in $\delta \mu$.
As to scale this out, \cref{fig:nernst} shows the rescaled Nernst coefficient $e_N \cdot \delta \mu^2/T^2$. This rescaled Nernst coefficient can be seen to exhibit two different scaling regimes (seen most clearly for high $T > G$ and $\delta \mu/G \geq 1$; panels (c) and (d) in \cref{fig:nernst}). The transition between the two regimes coincides with the onset of the $B$-linear regime in the longitudinal resistivity.
For lower $T< G$ (panel (a)) one is outside a %
perturbative regime: visually it appears one has reaches a novel scaling regime with the added observation that an increase of lattice amplitude now decreases the Nernst effect. This again matches the behavior of the longitudinal magnetoresistance (\cref{fig:magneto}(a)), which in this regime $T<G$ also becomes less sensitive to $B$ than in the approximate $B$-linear regime at higher $T/G$.

\section{Conclusion and outlook\label{sec:conclusion}}

Quantum critical systems can give rise to idiosyncratic transport signatures. This study of
DC and AC thermoelectric and magneto-transport in 2D quantum critical theories with strong translational symmetry breaking due to a spatially varying chemical potential lattice with zero average exhibits this again.
The most remarkable aspect is that the transport is never Drude-like in the sense that there is never a {\em single} longest-lived (momentum) relaxation mode controlling transport. In a 1D periodic potential electrical transport is purely quantum critical, whereas thermal transport is governed by two propagating modes originating in a collision of the would-be Drude mode with the Umklapp charge-diffusion mode. In a 2D lattice, on the other hand, the low-energy physics resembles that of Effective Medium Theory: currents follow meandering spatial paths of least resistance in a heterogeneous system built out of local regions that resemble alternating particle-filled and hole-filled states. This results in new weak diffusive modes that are not the global momentum-relaxation mode, but probably better seen as remnants of the local Drude modes from of the local charge-filled regions. This resemblance to results from Effective Medium Theory is also present in magneto-transport, where the longitudinal magnetoresistance shows a non-analytic unsaturating scaling behavior at large $B$ that is approximately linear. 

The context of this physics is familiar from studies of charge neutral graphene on various substrates with a periodicity larger than the intrinsic lattice. These studies have also shown the connection with Effective Medium Theory phenomenology \cite{cheianovRandomResistorNetwork2007a,rossiEffectiveMediumTheory2009,fernandesEffectiveMediumModel2023}. Our findings that one can have to the eye Drude-like transport co-existing with unsaturating $B$-linear-like magnetoresistance may be also relevant to the strange metal phase of high-$T_c$ cuprates. It provides a potential answer to the two-time scale puzzle needed to explain differing phenomenology of Hall and longitudinal transport, which has stubbornly eluded resolution since it was pointed out by Anderson.
A purely charge-neutral system as studied here will however not apply; for one, the Hall effect vanishes. More fundamentally, theoretical studies in SYK models have shown that one must have an IR fixed point determined by $SL(2,R)$ time-reparametrization invariance to explain local quantum criticality. In terms of its holographic dual, the $T=0$ near horizon geometry must be AdS$_2$, whereas charge-neutral systems have an AdS$_4$ geometry.
Once a finite density is added to the problem,
the question to answer is whether a low-temperature regime with AdS$_2$ physics might still give Effective Medium Theory multi-time-scale signatures, as long as $T/G \gtrsim 1$.\footnote{One can already deduce that for the characteristic temperature of a cuprate strange metal $T=100K$,
the regime $T/G = 1$ corresponds roughly to spatial variations on the order of \mbox{$500\,\mathrm{nm}$}
assuming the relevant speed is the Fermi velocity \mbox{$v_{\text{F}} \sim 5\cdot 10^{6}\,\mathrm{m/s}$}. 
The relevant modulation must therefore occur on scales far larger than the atomic lattice.}
A very recent SYK-based computation \cite{valentinisSuperlinearHallAngle2025} indeed finds different time-scales in ordinary and Hall transport, due to an effective increase of the Hall coefficient $R_H$ with $T$ --- a feature that also follows from a phenomenological match of cuprate DC transport to hydrodynamics \cite{amorettiHydrodynamicalDescriptionMagnetotransport2020}. 
As $R_H=B/n$ perturbatively, this naively implies zero-density at $T=0$, suggestively indicating that the fundamental physics is similar.

\acknowledgments

We thank Nicolas Chagnet for extensive discussions at the beginning of this project. E.N.\@ is funded by the Nano Area of Advance at Chalmers University of Technology. K.S.'s research is supported in part by Dutch Research Council (NWO) NWA Consortium {\em Emergence at all scales} (NWA.1630.23.007) and NWO Summit Grant {\em Quantum Limits} (SUMMIT.1.016). The numerical computations were carried out in part on the ALICE-cluster of Leiden University. We are grateful for their help. %

\appendix
\section{Numerical method}
\label{app:numerics}
The equations of motion were obtained in Mathematica with the help of the \textit{xAct} package \cite{brizuela_xpert_2009}.
To solve the resulting linear/non-linear equations of motion, we have used the %
\textit{PETSc} library in c \cite{petsc-efficient,petsc-user-ref}. For a guide on implementation of the latter we refer the reader to Ref.~\cite{bueler_petsc_2021}.

We have opted for a mix of pseudospectral (PS) and finite difference (FD) methods in the radial $z$-direction: we find an improved convergence at low temperatures/strong lattices with PS methods the at the cost of speed due to the global nature of the derivatives and use this to solve for the backgrounds. We do however find better numerical convergence with FD methods for the perturbations, along with a greatly appreciated speedup in the case of 2D lattices. In both cases we discretizes the $z$-direction on a Chebyshev-Lobatto grid,
\begin{equation}
    z_i = \frac{1}{2}\lsb 1 - \cos\lp\frac{ \pi n}{N}\rp \rsb, \quad n = 0, 1, \dots, N\,,
\end{equation}
which clusters points near the boundary and the horizon. In the transverse $xy$-directions, we used 8th/4th order finite differences on uniform grids for 1D/2D lattices.\footnote{In order to resolve the sharp Umklapp sound peak at large $T/G$ in the 2D AC response, we must increase the finite difference order in the $z$-direction to 8.} This leads to additional sparsity in the matrix discretizing the problem, and most importantly, allows for efficient partitioning of the matrix on multiple processors. PETSc is built for finite differencing, so to implement the radial pseudospectral derivatives we force PETSc not to split the domain along the $z$-direction across the processors.\footnote{This is achieved by first lying to PETSc's DMDA routine that the stencil width is that of the finite difference stencil and then explicitly telling PETSc how many non-zero entries per row there are (otherwise a lot of time is lost on memory allocation).}

To solve the non-linear equations for the backgrounds we used a GMRES non-linear solver, as the matrices are in general not symmetric due to the boundary conditions in the radial coordinate. To solve the linear equations (at each step of the non-linear algorithm, or for the linear equations of the perturbations), we used Schwarz domain decomposition implemented via an ASM preconditioner, distributing the calculation over 4--10 processors.
As a sub-preconditioner on each block, we used ILU or LU factorization with varying levels of overlap, depending on the numerical difficulty. The linear equations for the perturbations often require full LU factorization with 2 points of overlap, whereas it generally suffices with ILU(0) with 1 point of overlap for the non-linear background equations, expect for the very strongest lattices. The Stokes equations are easily solved with exact LU factorization on a single process. Note that since the Stokes equations \cref{eq:stokes} only depend on the derivatives of $p$ and $w$, one needs to fix a value for these functions at some point in the domain in order to remove a pair of zero modes of the matrix discretizing the equations.

\subsection{Numerical convergence}
The numerical difficulty increases as the temperature/lattice strength is reduced/increased. Throughout our calculations, we use units set by the horizon radius, which we fix to unity. In these units, the temperature is constant (for $B=0$), so lowering $T/G$ corresponds to increasing the lattice vector $G$. Maintaining a large ratio $\delta\mu/G$ therefore becomes increasingly challenging at low temperatures, since this requires $\delta\mu$ (in units of the horizon radius) to be correspondingly large.
\begin{figure}
    \centering
    \includegraphics{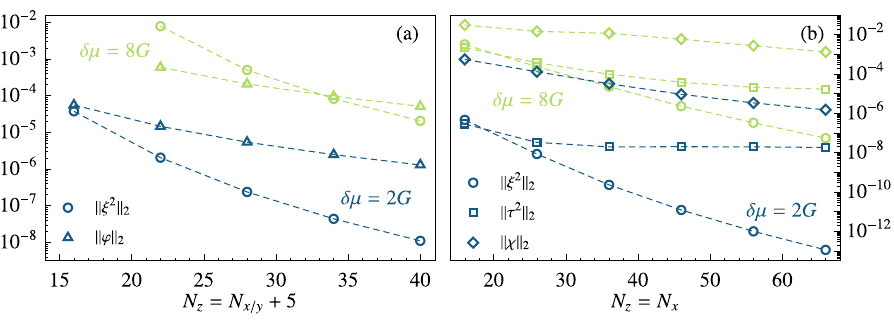}
    \caption{Convergence of the numerical algorithm. \textbf{(a)} Square of the de Turck vector $\xi^2$ and Maxwell gauge constraint scalar $\varphi$ in a 2D lattice for increasing grid sizes, where $N_x=N_y = N_z - 5$, at a low temperature $T = 0.1G$ and magnetic field $B=0.1G^2 = 10 T^2$. We find an exponential convergence with increased grid sizes. \textbf{(b)} Square of the de Donder vector $\tau^2$ and the Lorenz gauge scalar $\chi^2$, as well as the background de Turck vector, in a 1D lattice for increasing grid sizes. Here $N_x= N_z$, $T = 0.5G$ and $\omega=0.2G$, and the boundary conditions source a charge.}
    \label{fig:convergence}
\end{figure}

To quantify the convergence of our numerical algorithm, we monitor the norm of the constraints enforcing the gauge constraints: the de Turck vector $\xi^\mu$ is spacelike, so it suffices to check that $\xi^2 \ll 1$. In \cref{fig:convergence}(a) we show that the 2-norm $\| \xi^2\|_2$ converges exponentially with increased grid size, here using finite differences.\footnote{The 2-norm is taken over all grid points.}
The 2-norm of Maxwell gauge constraint scalar $\| \varphi\|_2$ (cf.~\cref{eq:bulk_EOM}) converges similarly.
In \cref{fig:convergence}(b), we illustrate the corresponding quantities for an AC calculation: the 2-norm of the square of the de Donder vector $\tau_\mu$ and Lorenz scalar $\chi = \nabla^\mu \delta A_\mu$. Here we computed the backgrounds using pseudospectral methods in the radial direction (due to the low temperature/strong lattices), but used finite differences for the perturbations. In general, the the perturbations exhibit weaker convergence with increased grid size. Both of these figures represent numerically challenging regions in the parameter space: at larger temperatures/weaker lattices, the convergence metrics look even better.

For challenging perturbation calculations at low temperatures (e.g., panels (a) and (d) in \cref{fig:AC1DallT}) for 1D lattices, we discretized the equations on a $80\times 60$ grid. For the backgrounds, this achieves a 2-norm of the de Turck vector ${\| \xi^2\|_2 < 10^{-4}}$ for the largest lattice amplitudes $\delta\mu/G = 6.0$, with an exponentially decreasing norm for weaker lattices. At high temperatures, for both 1D and 2D lattices, we are able to reach better convergence. Even on a moderately coarse $20\times 20\times 24$ grid, which we use for the AC response in \cref{fig:AC2dhighT}, ${\| \xi^2\|_2 < 10^{-6}}$ for $\delta\mu/G = 8.0$. For the Stokes calculations of the DC response we could be more liberal with the grid sizes and employed up to $45\times 45 \times 50$ grids %
where again $\| \xi^2\|_2 \lesssim 10^{-4}$ for the strongest lattices and lowest temperatures.

\bibliographystyle{JHEP}
\bibliography{biblio.bib}

\end{document}